\documentclass[twocolumn]{svjour3}
\usepackage{epsfig}
\usepackage{graphicx}
\usepackage{amsmath}
\usepackage{amssymb}
\usepackage{bbm}
\usepackage{diagbox}
\usepackage{tabularx}
\usepackage{xcolor}
\usepackage[ruled]{algorithm2e}
\usepackage{colortbl}
\usepackage{multirow}
\usepackage{bbding}
\usepackage{subfig}
\usepackage{ragged2e}
\usepackage{colortbl}
\usepackage{arydshln}
\usepackage{enumerate}
\usepackage{pifont}
\usepackage{makecell}
\usepackage{enumitem}
\usepackage{natbib}

\usepackage{hyperref}
\hypersetup{hidelinks,
	colorlinks=true,
	allcolors=black,
	pdfstartview=Fit,
	breaklinks=true}

\usepackage{xspace}
\makeatletter
\DeclareRobustCommand\onedot{\futurelet\@let@token\@onedot}
\def\@onedot{\ifx\@let@token.\else.\null\fi\xspace}

\definecolor{mygray}{gray}{.9}

\newcolumntype{P}[1]{>{\centering\arraybackslash}p{#1}}
\begin{document}

\title{Image Editing with Diffusion Models: A Survey}

\author{Jia Wang \and
    Jie Hu \and
    Xiaoqi Ma \and
    Hanghang Ma \and
    Xiaoming Wei \and
    Enhua Wu
}

\institute{
Jia Wang is with the University of Chinese Academy of Sciences, Beijing, China.
Jia Wang, Jie Hu, Xiaoqi Ma, Hanghang Ma and Xiaoming Wei are with Meituan, Beijing, China.
Jie Hu and Enhua Wu are with Key Laboratory of System Software (Chinese Academy of Sciences) and State Key Laboratory of Computer Science, Institute of Software, Chinese Academy of Sciences, Beijing, China.
Corresponding author: Jie Hu.
\email{wangj.infinite@gmail.com, hujie@ios.ac.cn}.
}

\maketitle

\begin{sloppypar}
\begin{abstract}
With deeper exploration of diffusion model, developments in the field of image generation have triggered a boom in image creation. As the quality of base-model generated images continues to improve, so does the demand for further application like image editing. In recent years, many remarkable works are realizing a wide variety of editing effects. However, the wide variety of editing types and diverse editing approaches have made it difficult for researchers to establish a comprehensive view of the development of this field. In this survey, we summarize the image editing field from four aspects: tasks definition, methods classification, results evaluation and editing datasets. 
First, we provide a definition of image editing, which in turn leads to a variety of editing task forms from the perspective of operation parts and manipulation actions. Subsequently, we categorize and summary methods for implementing editing into three categories: inversion-based, fine-tuning-based and adapter-based. In addition, we organize the currently used metrics, available datasets and corresponding construction methods. At the end, we present some visions for the future development of the image editing field based on the previous summaries.

\end{abstract}

\section{Introduction}
\label{sec:Introduction}

\begin{figure*}[t]
\centering
\includegraphics[width=1\textwidth, trim=0cm 0cm 2cm 0cm, clip]{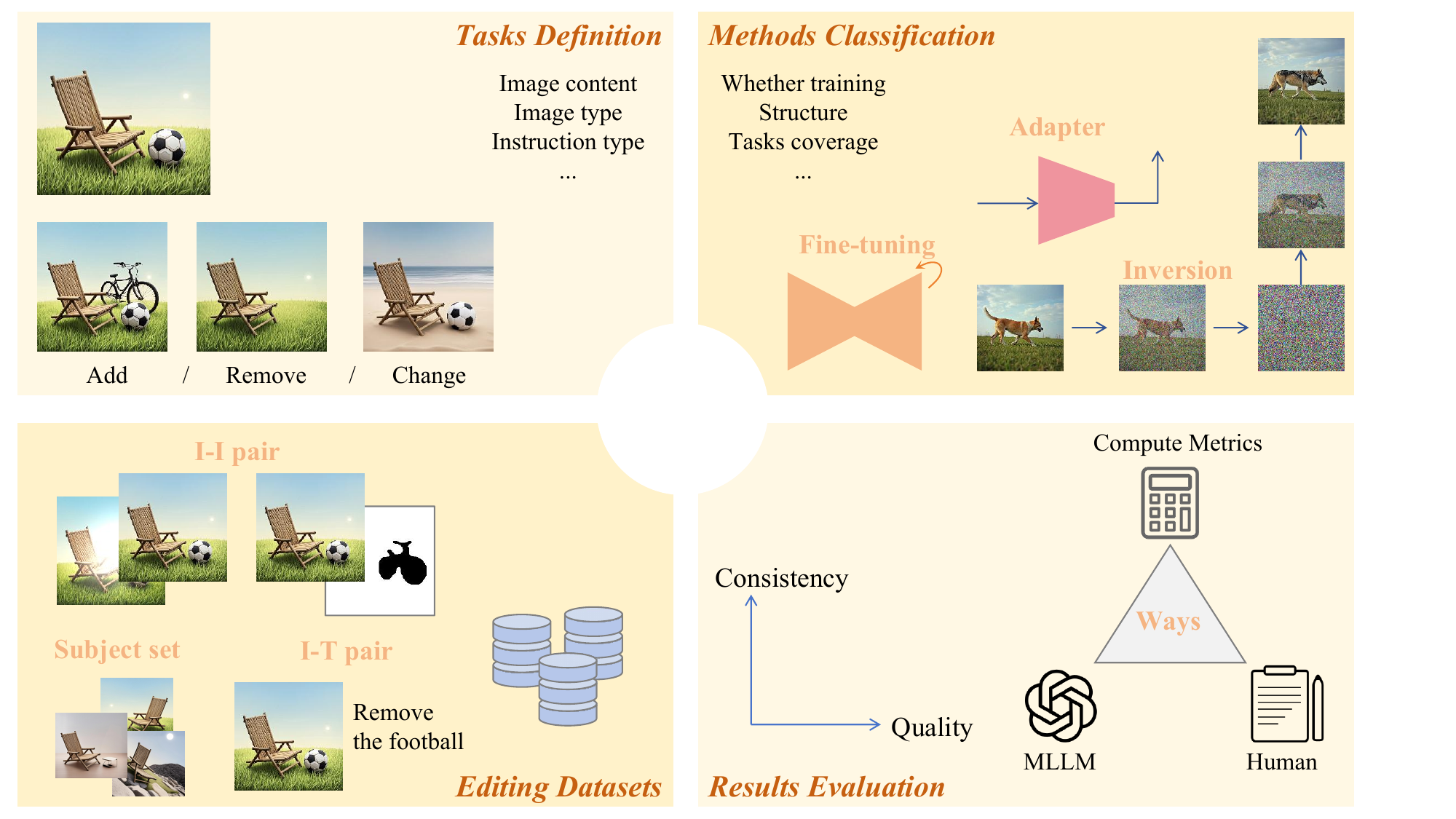} 
\caption{An overview of our survey, which includes four main parts: editing tasks, methods classification, results evaluation and editing datasets.}
\label{fig:oveview of survey}
\end{figure*}

With the continuous advancement in image generation technology, current text-to-image models~\citep{nichol2021glide, saharia2022photorealistic, ramesh2022hierarchical, rombach2022high, podell2023sdxl} like Stable Diffusion (SD)~\citep{rombach2022high} are capable of producing high-resolution, richly detailed images. Although these models can use text prompts to convey a general concept of an image, rendering more intricate details still necessitates the input of raw images or more precise representations. Moreover, numerous practical applications, such as retouching, backgrounds altering, and virtual try-ons~\citep{choi2021viton, choi2024improving, morelli2023ladi, li2024unihuman, zeng2024cat}, typically require modifications to existing images, wherein most of the input image content is preserved with only specific areas altered. Text-to-image generation models generally lack this degree of precise manipulation, highlighting the importance of image editing technology for these applications. Generative Adversarial Networks (GANs)~\citep{goodfellow2014generative, karras2019style, mirza2014conditional, brock2018large}, as pioneering generative models, have significantly advanced image editing~\citep{isola2017image, zhu2017unpaired}, representing a shift from solely generation-focused tasks to editing-oriented researches.

Recently, the emergence of diffusion models has significantly contributed to the explosion of AI-generated content (AIGC)~\citep{esser2021taming, zhao2016energy, rezende2015variational, van2016pixel, papamakarios2017masked, ramesh2021zero}. Diffusion models~\citep{ho2020denoising, song2020denoising, song2019generative, song2020score, song2020improved, bao2022analytic} define the image generation process as learning a mapping from a simple distribution, specifically a Gaussian distribution, to the distribution of natural images. Inspired by non-equilibrium thermodynamics~\citep{sohl2015deep}, they divide this process into multiple steps, determining each step's path by incrementally adding noise. Unlike Variational Autoencoders (VAEs)~\citep{kingma2013auto, higgins2017beta, van2017neural, razavi2019generating} and Generative Adversarial Networks (GANs), this multi-step operation of diffusion models enables greater computational depth with equivalent model parameters by cyclically inputting noise images into it. As image information is decomposed across several denoising steps, each step is tasked with learning reduced content, thus easing complexity and significantly enhancing the quality ceiling for generated images. Due to their advanced modeling capabilities, diffusion models have seen applications across various fields, including image generation~\citep{ramesh2022hierarchical, rombach2022high, saharia2022photorealistic, ho2022classifier, ho2022cascaded, dhariwal2021diffusion, meng2023distillation, bansal2024cold, phung2023wavelet}, video generation~\citep{ho2022video, singer2022make, ho2022imagen, blattmann2023align, wu2023tune, ge2023preserve, zhou2022magicvideo}, audio generation~\citep{kong2020diffwave, liu2023audioldm, gong2022diffuseq, ruan2023mm}, image restoration~\citep{saharia2022image, ozdenizci2023restoring, shang2024resdiff, gao2023implicit, guo2023shadowdiffusion, xia2023diffir} and image editing.

Numerous studies have employed diffusion models for image editing tasks. However, the broad definition of editing encompasses various task forms, which may differ across studies, posing challenges for researchers seeking to build upon existing work. On the one hand, the diversity of implementation methods makes it difficult for researchers to extract core structures and algorithms from an extensive body of literature. On the other hand, the variety of tasks makes it harder to find a unified and relevant framework for image editing. Consequently, a comprehensive survey of image editing with diffusion models is essential and valuable.

While some existing surveys describe diffusion models thoroughly, they often do not focus specifically on image editing tasks~\citep{xing2024survey, yang2023diffusion, cao2024survey, croitoru2023diffusion, po2024state}. Other reviews address controllable image generation or image editing, concentrating on various implementation approaches and predominantly introducing and categorizing editing methods~\citep{cao2024controllable, huang2024diffusion, shuai2024survey, zhan2024conditional}. Despite covering a substantial amount of related work, these reviews generally lack detailed analysis from other perspectives. Given the diverse nature of image editing, it is necessary to organize editing tasks in an more comprehensive way. Simultaneously, the categorization of editing methods should avoid excessive complexity, instead highlighting the similarities and differentiating features of these methods from a broader perspective. Furthermore, with the continual advancements in computing power and model size, it is essential to organize existing image editing datasets and their construction methodologies to facilitate further research.

\begin{figure*}[t]
\centering
\includegraphics[width=1.0\textwidth, trim=0cm 0cm 0cm 0cm, clip]{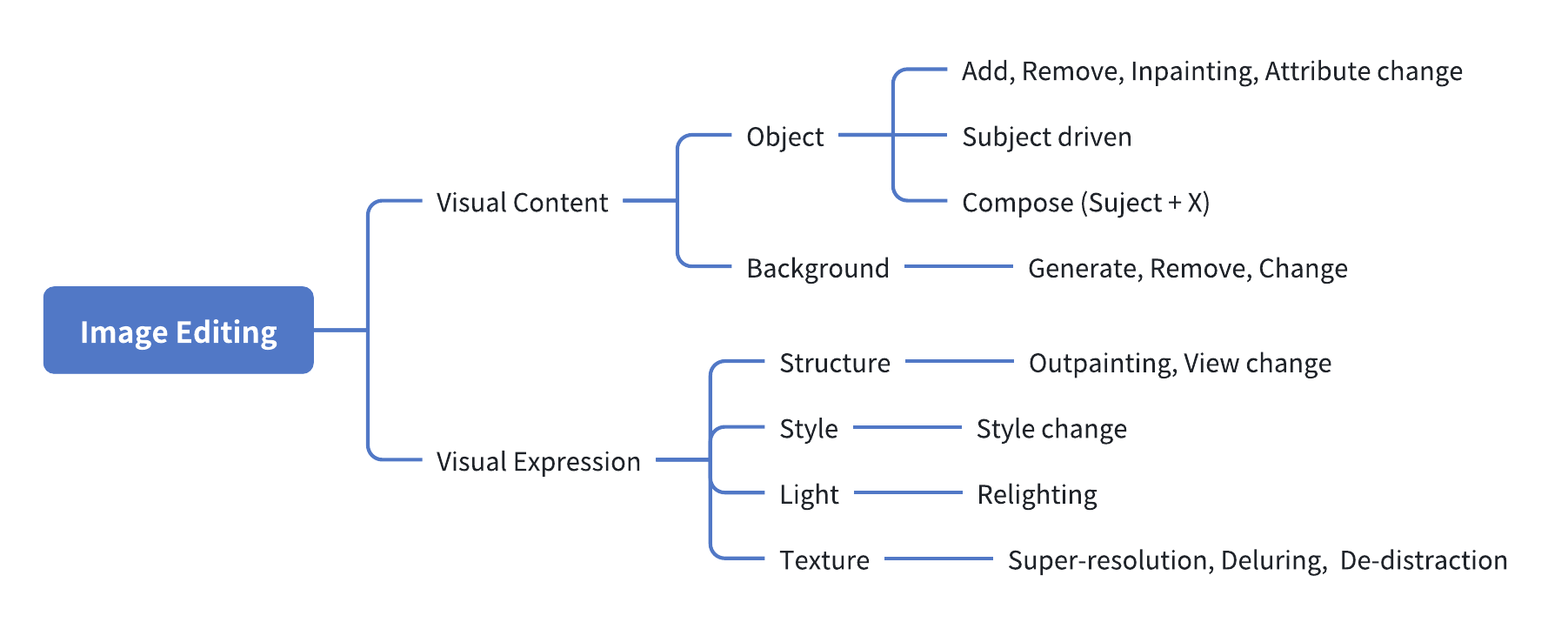} 
\caption{Partition of images and corresponding editing tasks. An image can be divided into two primary components: visual content and visual expression, each of which can be further segmented into more detailed concepts. Each editing task can be regarded as the manipulation of these underlying concepts.}
\label{fig:Editing Tasks Map}
\end{figure*}

In this work, we conduct a systematic review and summarization of editing tasks, editing methods, result evaluation, and editing datasets, as shown in Fig.~\ref{fig:oveview of survey}. Initially, we define image editing from the perspectives of operation components, manipulation actions, and instruction types. We categorize editing objects (images) into two primary components: visual content and visual expression, providing a detailed overview of each part. Different editing tasks are distinguished by variations in operational components and manipulation actions. To enhance comprehension of task formats, we offer visual examples for prime editing tasks. 
Regarding the classification of editing methods, we divide them into three categories based on the degree of modification of the base model: inversion-based methods (requiring no modification to the base model and minimal additional parameters), fine-tuning-based methods (primarily altering base model parameters), and adapter-based methods (requiring no modification to base model parameters but necessitating adapter training). Subsequently, we compile current evaluation metrics for image editing and describe the process of designing related benchmarks. We then summarize existing image editing datasets, detailing the steps involved in their construction. Finally, we highlight several existing challenges in the field of image editing and introduce potential future research directions.

Overall, we organize and summarize various dimensions of image editing, proposing new inductive perspectives. The primary distinction between this work and previous surveys lies in our comprehensive observations and clearer, simpler classifications. In Section~\ref{sec:Tasks Definition}, we define the tasks and formats of image editing. Section~\ref{sec:Editing Methods} classifies existing editing models. Section~\ref{sec:Evaluation} presents related evaluation metrics and the construction of editing benchmarks. Section~\ref{sec:Datasets} organizes existing datasets and their construction methods. In Section~\ref{sec:Challenges and Future Directions}, we explore the challenges and future directions of image editing. Finally, in Section~\ref{sec:Conclusion}, we summarize our work.

\section{Tasks Definition}
\label{sec:Tasks Definition}

\begin{figure}[t]
\centering
\includegraphics[width=0.5\textwidth, trim=0cm 12cm 18cm 0cm, clip]{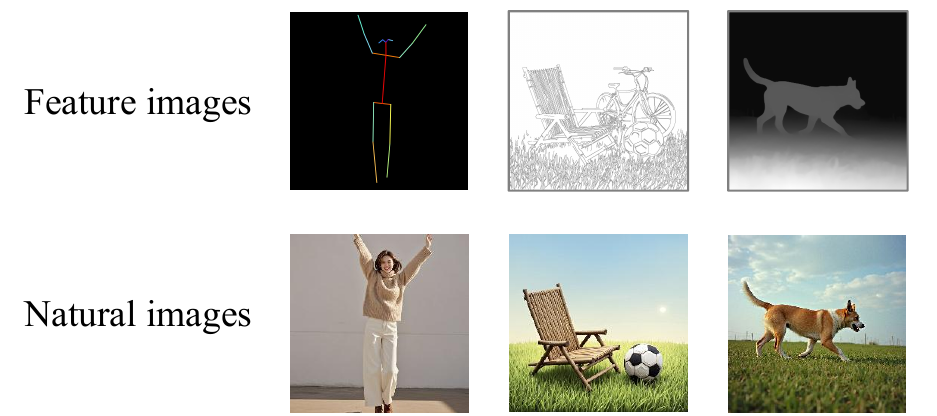} 
\caption{Examples of feature images and natural images.}
\label{fig:Editing Image Definition}
\end{figure}

\begin{figure}[t]
\centering
\includegraphics[width=0.5\textwidth, trim=0cm 4cm 21cm 0cm, clip]{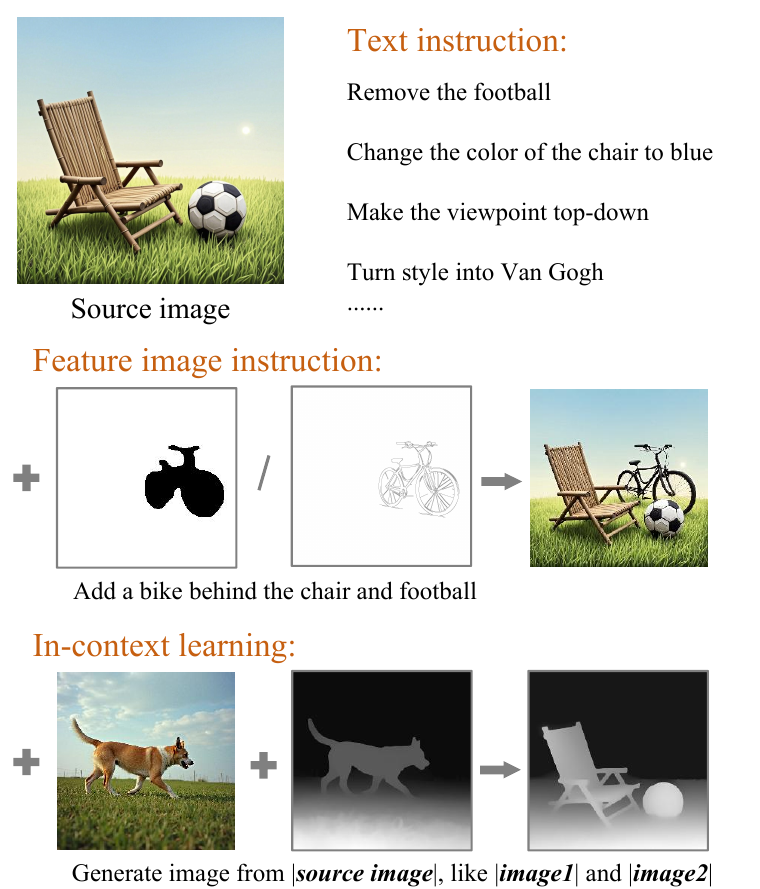} 
\caption{Different instruction types. Text instructions are suitable for general editing scenarios. Feature image instructions enable more fine-grained editing. In-context learning can convey different editing modes.}
\label{fig:Instruction types}
\end{figure}

Current research efforts have established various classifications for image editing tasks. In these surveys, \citet{shuai2024survey} propose a hierarchical classification system distinguishing between content-aware editing and content-free editing. The former is further subdivided into local editing (encompassing object manipulation, attribute manipulation, spatial transformation, and inpainting) and global editing (including style change and image translation). Content-free editing comprises subject-driven customization and attribute-driven customization. While this framework initially distinguishes tasks based on content involvement, the classification exhibits structural imbalance. Customization emerges as the sole distinct category at the primary level, and subsequent local/global partitioning leads to disproportionate task allocation favoring local editing.

In contrast, \citet{huang2024diffusion} present a tripartite classification, categorizing editing operations as: (1) semantic editing (object addition/removal/replacement, background modification, emotional expression alteration), (2) stylistic editing (color adjustment, texture modification, holistic style transformation), and (3) structural editing (object relocation, dimensional/shape alteration, pose/action modification, perspective adjustment). Although this three-category system achieves better task distribution equilibrium, it primarily adopts a bottom-up approach that aggregates existing tasks through similarity clustering until full coverage is attained. Our analysis suggests the need for a more principled taxonomy that systematically emerges from fundamental editing objectives rather than organizing established task types.

\begin{figure*}[ht]
\centering
\includegraphics[width=1\textwidth, trim=0cm 7cm 0cm 0.3cm, clip]{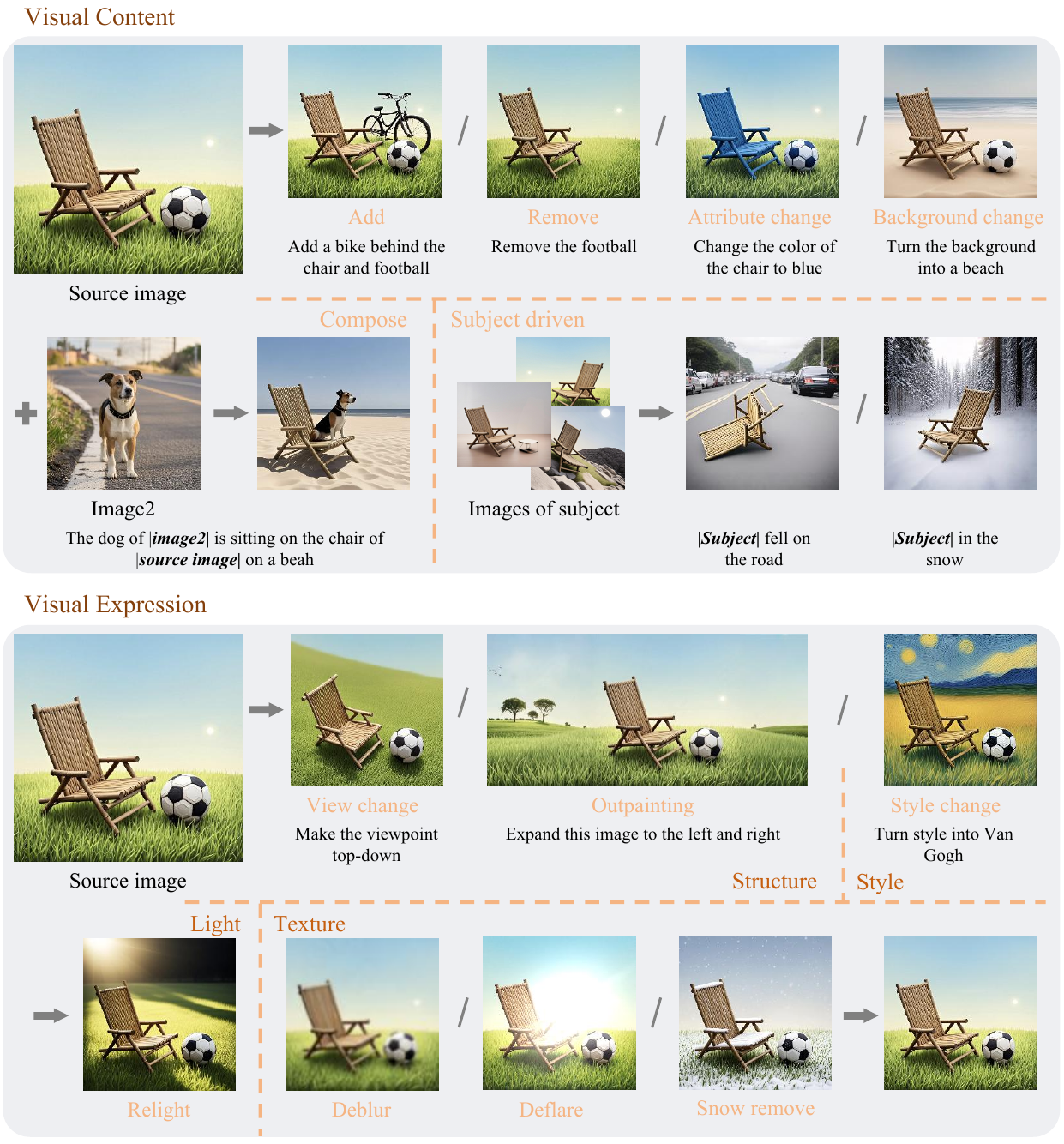} 
\caption{Examples of different editing tasks. From the perspective of simplicity, we standardized the source images for most editing examples.}
\label{fig:Editing Tasks Visualization}
\end{figure*}

\subsection{Definition of Editing}
\label{sec:Definition of Editing}

Before listing image editing tasks, it is crucial to establish a precise definition of image editing. We define image editing as the modification of existing images to meet user-specified intentions. Based on this definition, there are four key issues to consider: (1) what type of image to modify, (2) which parts of the image to modify, (3) what kinds of modifications to apply, and (4) through what form to express editing instructions?

Our taxonomy first categorizes images into feature images and natural images according to the degree of structural detail (as shown in Fig.~\ref{fig:Editing Image Definition}). Feature images are technical representations, such as keypoint images, edge images, and depth images. These images contain sparse structural information and are primarily used in specialized analytical applications. Conversely, natural images exhibit rich textural and chromatic detail, closely resembling human visual perception. For example, they can be obtained through photographic capture, hand-painting, or generative synthesis. This distinction helps differentiate natural images, which are intuitive for non-specialists, from feature images, which primarily serve technical workflows. While our research focuses on natural image editing, this classification is valuable for understanding broader visual computing paradigms. Within this framework, feature-to-natural image transformations represent controllable generation processes, while natural-to-feature image conversions encompass established computer vision processes like object detection and feature extraction. The proposed taxonomy not only clarifies image editing operations but also establishes a structured framework for analyzing related visual computing tasks.

We further suggest dividing image components into two categories: visual content and visual expression. Visual content comprises concrete image elements, including objects and backgrounds, representing objectively identifiable items. Visual expression encompasses elements such as structure, style, light, and texture, representing higher-level content that is closely related to human aesthetic perception. This division aids in a comprehensive understanding of the structural composition of images and clarifies which parts are targeted by editing tasks.

Regarding user operations, we initially considered basic actions such as add, delete, and combine. Upon further reflection, we propose that editing operations can be more broadly defined as add, delete, change, and combine. The change operation serves as a comprehensive descriptor suitable for modifications to any image content, while add and delete are more appropriate for operations on specific objects or backgrounds. The combine operation involves merging two or more image components, typically involving at least one object. This approach to defining tasks by combining actions and objects demonstrates both simplicity and generalizability.

How users convey their editing instructions is another area that warrants discussion. One of the most basic methods is through textual descriptions, such as “Remove object A from the picture.” However, textual descriptions might lack the precision necessary to adequately convey structural or detailed information. In contrast, the image itself serves as an effective medium for expressing instructions. Users can provide control images, which can be either feature images or natural images, to express their editing intentions. In addition, images can convey not only descriptive information but also the mode of editing. Typically, users can input two images, one depicting the pre-edit state and the other the post-edit state, enabling the model to understand the editing pattern and subsequently apply the learned editing pattern to other images. This approach, known as in-context learning, enhances the accuracy and efficiency of editing and provides users with a more intuitive way to edit. Besides these, drag-based operations can also serve as instructions~\citep{shi2024dragdiffusion, zhao2024fastdrag}. As illustrated in Fig.~\ref{fig:Instruction types}, this multimodal instruction framework combines the expressiveness of visual inputs with the flexibility of textual input, effectively bridging the intention-specification gap in complex editing tasks.

\subsection{Editing Tasks}
\label{sec:Editing Tasks}

To better demonstrate the implementation of various image editing tasks, it is essential to visualize input-output examples. Fig.~\ref{fig:Editing Tasks Visualization} illustrates multiple image editing tasks, categorized into two primary dimensions: visual content editing and visual expression editing.

In the domain of visual content editing, the system supports a wide range of functionalities, encompassing both basic and advanced operations. Basic tasks include object addition or removal (object add/remove), color modification (attribute change), and background replacement (background change). Advanced tasks, such as composition (composing), enable the fusion of elements from different images by integrating reference images. For example, this may involve combining a chair from the original image and a dog from an external image, then placing them in a target scene such as a beach, thereby overcoming the limitations of single-image editing. In addition, subject-driven tasks require multiple photographs of a subject captured in various environments. The model learns the subject's structure and details from these inputs and subsequently generates new images of the subject in different contexts based on textual instructions. As depicted in the figure, a chair shown in scenes such as grass, indoors, and on rocks serves as input, and through textual directives, the system generates images of the chair placed on a road and in the snow.

For visual expression editing, we categorize tasks into four components: structure, style, lighting, and texture. Each component has specific editing operations. Structural editing involves changing the composition of the image, such as view change, outpainting, and layout editing~\citep{xia2025consistent}. These tasks may transform the perspective or extend the image horizontally. Style change focuses on altering the artistic appearance, such as converting the image into a Van Gogh-style artwork. Relighting modifies light effects, such as creating a directional effect from left to right. Texture-related tasks modify surface details and clarity. Examples include deblurring, deflaring, and snow removal.

\section{Editing Methods}
\label{sec:Editing Methods}

The denoising network is a pivotal component of the diffusion model, iteratively processing noise to generate the final latent representation. Image editing techniques that utilize diffusion models typically employ a pre-trained text-to-image model, incorporating specific modifications to execute editing tasks. These modifications can encompass alterations to noise latents, attention maps, network architectures, and network parameters. We categorize existing editing approaches into three primary classes based on the extent of modifications applied to the base model. First, inversion-based methods achieve editing by adjusting key intermediate variables in the denoising process, such as noise latents and attention maps, without modifying the base model's parameters. Second, fine-tuning-based methods perform editing by introducing an interface for the original image input into the base model, and subsequently fine-tuning it with a limited set of parameters. Third, adapter-based methods achieve editing by using various adapters to process multimodal editing prompts and incorporate this information into the base model without altering its parameters.

\begin{figure*}[t]
\centering
\includegraphics[width=1\textwidth, trim=0cm 2cm 2cm 0cm, clip]{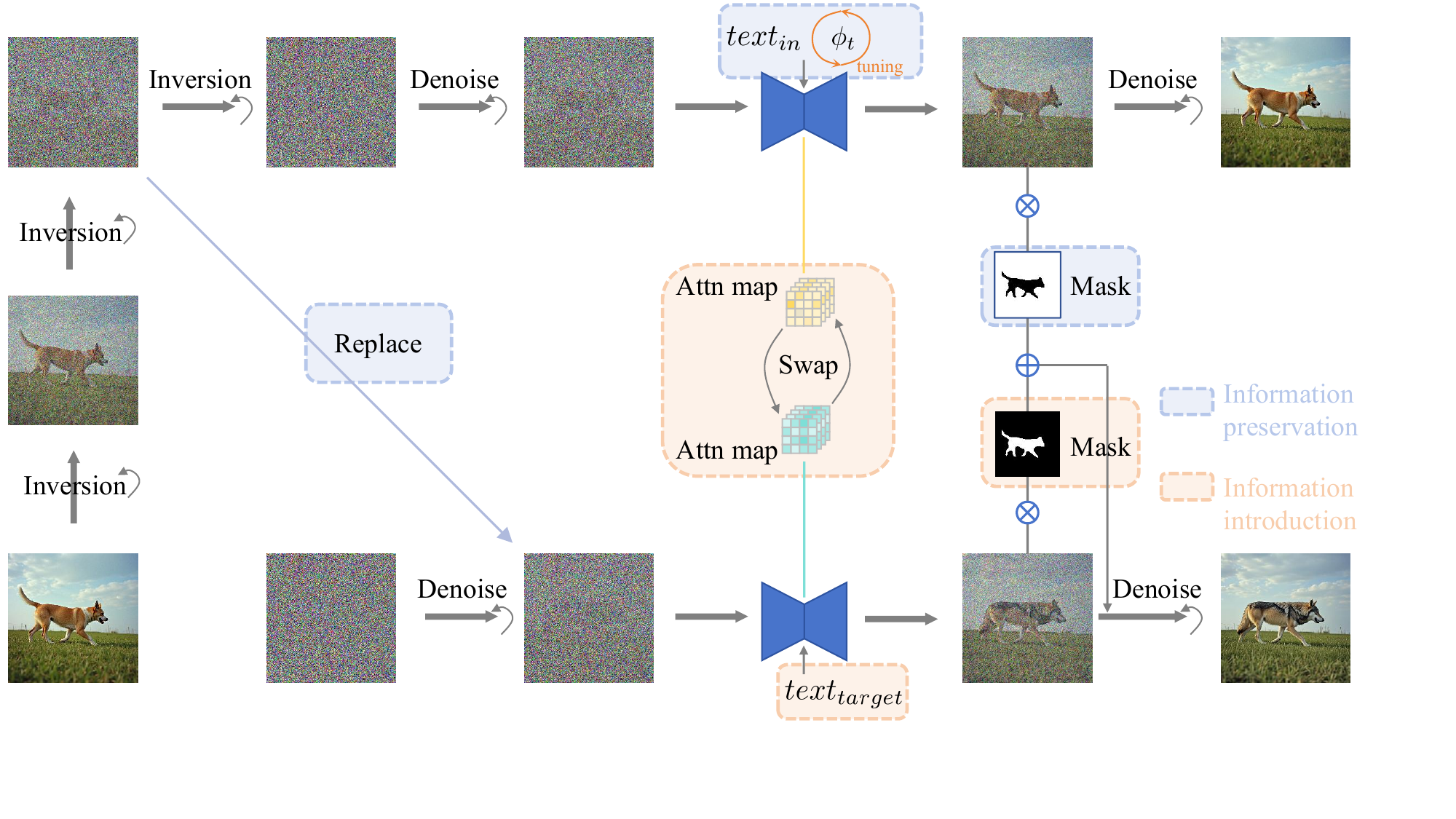} 
\caption{An overview of inversion-based methods. We select four representative methods: SDEdit~\citep{meng2021sdedit}, Null-text inversion~\citep{mokady2023null}, Prompt-to-Prompt~\citep{hertz2022prompt}, and Blended Diffusion~\citep{avrahami2022blended}, and demonstrate their core mechanisms.} 
\label{fig:inversion-based}
\end{figure*}

\subsection{Inversion-Based Methods}
\label{sec:Inversion-Based Methods}

Inversion-based approaches perform image editing primarily by altering the intermediate noise latents and attention maps within the denoising process of diffusion models. These methods encompass two critical considerations: preservation of the original image information and precise modification of the intended editing targets. Preserving the original image's integrity ensures that non-targeted elements remain consistent, thereby maintaining the overall coherence and quality of the output. Simultaneously, modifying the editing targets allows for specific alterations as determined by user instructions or predefined objectives, enabling controlled image transformations. By carefully balancing these aspects, inversion-based methods ensure that essential image details are retained while allowing for targeted modifications. To elucidate these two fundamental components—preservation and modification—we have selected several representative studies, as shown in Fig.~\ref{fig:inversion-based}. Through the analysis of these exemplary methods, we aim to highlight the strengths and limitations of the inversion-based approach, thereby providing a comprehensive understanding of its application in contemporary image editing tasks.

\subsubsection{Information Preservation}
For information preservation, a notable example of preserving original information is SDEdit~\citep{meng2021sdedit}. SDEdit partially distorts the original image information by inverting within the range of $t \in [0.3, 0.6]$, thereby retaining some original image details. By subsequently denoising the inverted image with the same steps, the edited image is generated. Another significant contribution in this area is Null-text inversion~\citep{mokady2023null}. Due to the stochastic nature of the diffusion process, achieving reversibility is often challenging. While some methods employ DDIM inversion~\citep{song2020denoising}, they frequently struggle to accurately restore details. To improve editing precision, null-text inversion replaces the null-text embedding at each denoising step with learnable parameters and uses the mean squared error (MSE) between the original intermediate noise latent and the latent noised once more (followed by restoration) as a loss function. This approach enables more precise restoration of the original image.

Many studies have further optimized Null-text inversion. PTI~\citep{dong2023prompt} introduces a novel inversion method that rapidly and accurately reconstructs the original image by optimizing a learnable conditional embedding, thereby providing a robust foundation for sampling high-fidelity edited images. Notably, this method does not require user-provided masks or textual descriptions of the input image, enhancing user-friendliness and enabling the handling of real-world images across various domains. NPI~\citep{miyake2023negative} achieves comparable reconstruction accuracy through forward propagation rather than an optimization-based process, substantially accelerating the editing workflow. This approach minimizes computational time without necessitating complex optimizations, thus improving overall editing efficiency. Additionally, EDICT~\citep{wallace2023edict} accomplishes mathematically precise DDIM inversion of real images by maintaining and alternately utilizing two coupled noise vectors. Although this technique effectively doubles the computational time of the diffusion process, it significantly enhances the accuracy of image reconstruction, particularly for complex images.

Besides these, several studies have extended beyond DDIM methodologies. CycleDiffusion~\citep{wu2023latent} proposes a DDPM inversion technique that recovers a sequence of noise vectors generated during the DDPM sampling process, enabling the perfect reconstruction of images. In contrast to DDIM, DDPM involves T+1 noise vectors in the generation process, each matching the dimensionality of the output. Furthermore, RF inversion~\citep{rout2024semantic} introduces an image inversion and editing method based on rectified stochastic differential equations. The core concept involves employing dynamic optimal control, derived from a linear-quadratic regulator, to achieve the inversion of rectified flows (RFs). This approach offers a sophisticated framework for precise image manipulation and reconstruction, expanding the capabilities of current image editing methodologies.

\subsubsection{Information Introduction}

Information can be introduced by manipulating noise latent variables or attention maps during the denoising process. These techniques facilitate efficient image editing through various technical approaches.

Prompt-to-Prompt (P2P)~\citep{hertz2022prompt} edits images by altering the attention maps, conceptualizing editing as the addition, deletion, or modification of terms in the text prompt. With a one-to-one correspondence between text tokens and attention maps, P2P accomplishes editing by adjusting slices of the attention maps during generation. It introduces a re-weighting strategy to control the intensity of edits by multiplying newly added attention maps slices by a scale factor; a factor greater than one indicates stronger changes, while a factor less than one suggests changes in the opposite direction.
Blended Diffusion~\citep{avrahami2022blended} addresses the editing process by employing a mask to restrict modifications to specific regions of the original map, thereby directing the model’s focus to areas requiring changes. Since this mask is applied at every denoising step, it serves as a robust anchor, guiding noise generation and ensuring consistent edges within the masked regions.

In contrast to P2P, Pix2Pix-Zero~\citep{parmar2023zero} does not require user-defined text prompts. Instead, it automatically identifies editing directions within the text embedding space to perform modifications. This approach analyzes the editing directions in the text embedding space and selects regions that necessitate alteration, thereby enabling zero-shot image editing without the need for explicit prompts.
Plug-and-Play (PnP)~\citep{tumanyan2023plug} achieves more fine-grained editing effects by directly inserting the spatial features from the reconstruction branch and the query and key matrices from self-attention into the editing branch. Unlike P2P, which replaces attention maps, PnP corrects deviation paths by directly adding differences, thereby enhancing both editing efficiency and effectiveness. This method can be implemented with merely three lines of code, significantly reducing computation time and the necessity for extensive optimization processes.

Blended Latent Diffusion~\citep{avrahami2023blended} extends the Blended Diffusion technique by applying it within the latent space of Latent Diffusion Models (LDMs), thereby further improving editing efficiency through more effective local modifications. 
Differential Diffusion~\citep{levin2023differential} employs Gaussian blur to process the mask image, achieving a more natural transition effect between edited and original regions. This approach mitigates unnatural appearances at the boundaries of edited areas, enhancing the overall naturalness and visual coherence of the editing outcome.
MasaCtrl~\citep{cao2023masactrl} introduces a training-free image synthesis and editing method that enhances consistency by replacing the Key and Value components in self-attention with those in mutual self-attention. Unlike P2P, which directly replaces attention maps, MasaCtrl utilizes mutual attention control to integrate the source image content with layouts synthesized from text prompts and other controls. This strategy prevents attention interference between foreground and background elements by using cross-attention maps as masks, thereby maintaining the integrity of both regions during the editing process.

\subsection{Fine-Tuning-Based Methods}
\label{sec:Fine-tuning-Based Methods}

Fine-tuning-based approaches involve modifying the entire base model to adapt it to editing tasks. These methods can be broadly categorized into training time fine-tuning and testing time fine-tuning. Training time fine-tuning refers to the process of adjusting the base model to develop a generalizable editing model. This model is capable of handling various editing tasks within a single inference after being fine-tuned. Due to their emphasis on generalizability, training-time fine-tuning methods typically require the preparation of extensive training data and incur significant computational costs during the training phase. The advantage of this approach lies in its ability to perform multiple editing tasks efficiently once the model has been fine-tuned. 
Test time fine-tuning, on the other hand, involves adjusting the model parameters specifically for particular datasets or individual editing tasks at inference time. This method is computationally more efficient when the number of editing tasks is limited, as it avoids the need for extensive retraining. However, when confronted with a large number of diverse editing tasks, training-time fine-tuning methods become more advantageous due to their ability to generalize across different tasks without requiring repeated adjustments.

\begin{figure}[t]
\centering
\includegraphics[width=0.5\textwidth, trim=0cm 7.5cm 15cm 0cm, clip]{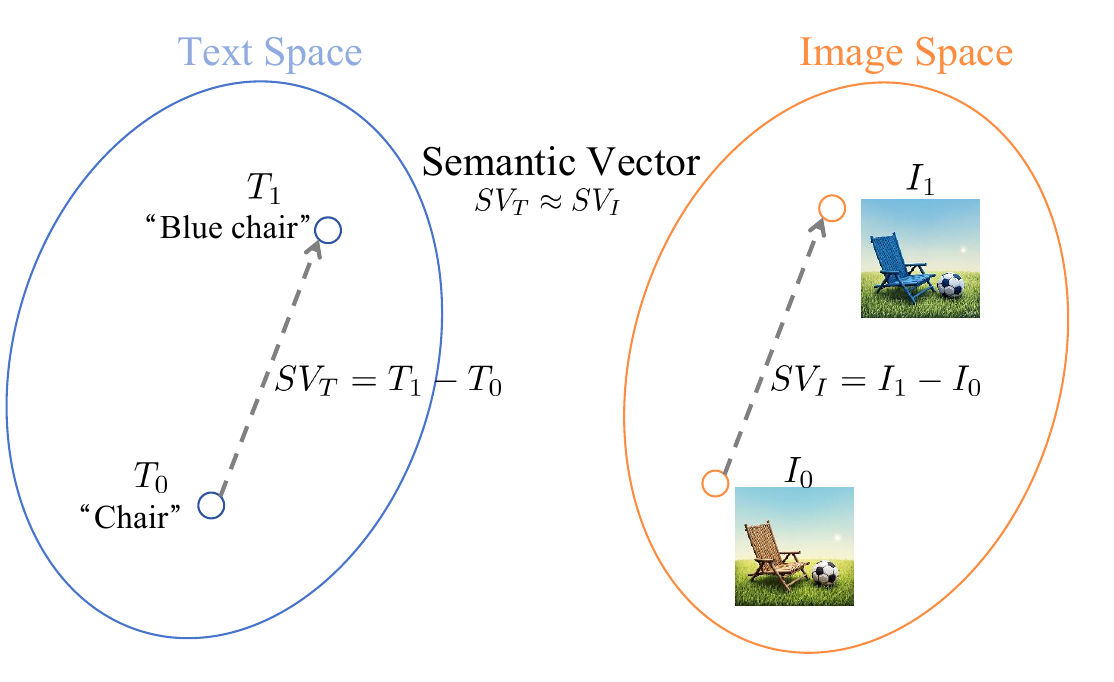} 
\caption{Sematic vector alignment in text and image space.}
\label{fig:fine-tuning-based}
\end{figure}

\subsubsection{Training Time Fine-tuning}
\label{sec:Training Time Fine-tuning}

A seminal work on training time fine-tuning is InstructPix2Pix~\citep{brooks2023instructpix2pix}. This method initially collects a substantial number of edited image pairs using Prompt-to-Prompt (P2P) as training data, with a detailed data construction process elaborated in Section~\ref{sec:Datasets}. Subsequently, these paired data are utilized to fine-tune the base model. However, many base models lack an interface for the input of the original image. To address this, InstructPix2Pix expands the number of convolutional channels, thereby enabling the splicing of the image input into the first convolutional layer. Additionally, it introduces a multi-conditional control mechanism:
\begin{equation}
\label{eqn:text and image class guidance}
\begin{aligned}
\tilde{f_{\theta}}\left(z_{t}, c_{I}, c_{T}\right)= & f_{\theta}\left(z_{t}, \varnothing, \varnothing\right) \\
& +s_{I} \cdot\left(f_{\theta}\left(z_{t}, c_{I}, \varnothing\right)-f_{\theta}\left(z_{t}, \varnothing, \varnothing\right)\right) \\
& +s_{T} \cdot\left(f_{\theta}\left(z_{t}, c_{I}, c_{T}\right)-f_{\theta}\left(z_{t}, c_{I}, \varnothing\right)\right), 
\end{aligned}
\end{equation}
where $f_{\theta}$ and $z_t$ stands the diffusion network and noise latent, $c_I$, $c_T$, $s_I$ and $s_T$ stand the reference image, reference text and their strength. This equation demonstrates the process of isolating pure control information. It does so by subtracting the portion obtained from conditional control from the portion associated with unconditional control. By appropriately scaling and summing different control information, varying control intensities and common control can be achieved.

Similarly, emu-edit~\citep{sheynin2024emu} employs a comparable approach to inject the original image. The key distinction lies in emu-edit is the categorization of different editing tasks and the establishment of a learnable task embedding to differentiate these tasks. This enables the model to segregate various editing tasks, facilitating more precise control. Structurally, emu-edit integrates the task embedding into the U-Net architecture through cross-attention mechanisms and incorporates it into the timestep embedding. To ensure that related editing tasks reinforce each other, MoEController~\citep{li2023moecontroller} incorporates a Mixture-of-Experts (MoE) architecture with three specialized experts dedicated to fine-grained local translation, global style transfer, and complex local editing tasks.

The above mentioned methods operate on the U-Net structure, while a representative approach for the latest DiT-based pedestal model is Omini-control~\citep{tan2024ominicontrol}. Unlike previous methods, Omini-control ingeniously splices the processed original image token with the input token of the DiT Block, achieving the injection of the original image with only a marginal increase in the number of parameters. 
UNO~\citep{wu2025less} proposes Universal rotary position embedding(UnoPE), and extends single-graph reference to multi-graph reference. It uses a multi-stage training strategy that gradually increases the number of reference graphs, making the learning process more stable.

Furthermore, several other notable works contribute to enhancing training-time fine-tuning. \citet{chakrabarty2023learning} improve dataset quality and enhance supervisory signals to enable models to execute object-based image editing instructions more accurately. InstructAny2Pix~\citep{li2023instructany2pix} introduces a multimodal instruction-following image editing system capable of flexibly editing input images based on complex instructions that integrate audio, images, and text. Additionally, HIVE~\citep{zhang2024hive} integrates Reinforcement Learning from Human Feedback (RLHF) into instructional image editing, further enhancing the accuracy and relevance of the editing outcomes.

\subsubsection{Testing Time Fine-tuning}
\label{sec:Testing Time Fine-tuning}

Testing time fine-tuning is a pivotal approach for realizing subject-driven tasks in image editing. Textual Inversion~\citep{gal2022image} employs an infrequently used word to represent an arbitrary object, image, or concept. Initially, the word is mapped to a learnable embedding, which is then trained using 3 to 5 user-provided photos to accurately encapsulate the desired concept. Once the embedding is learned, users can combine it with other textual prompts to generate a variety of images. For instance, by providing 3 to 5 photos of a chair, the model can produce images of the chair in various unseen scenes, as shown in Fig.~\ref{fig:Editing Tasks Visualization}.

Another notable method in subject-driven tasks is DreamBooth~\citep{ruiz2023dreambooth}. Like Textual Inversion, DreamBooth utilizes a special token to represent a particular subject. However, DreamBooth fine-tunes the parameters of the entire denoising network, enabling more nuanced consistency in the generated images. This comprehensive modification increases the likelihood of maintaining detailed and accurate representations of the subject. Nevertheless, altering the entire denoising network can lead to knowledge forgetting, where the model may lose the characteristics of other classes. To mitigate this issue, DreamBooth introduces a Class-specific Prior Preservation Loss, which helps preserve the class-specific features of the subject while allowing targeted edits.

Other test-time fine-tuning methods guide the transition by utilizing the alignment between text and images in the CLIP space, as shown in Fig.~\ref{fig:fine-tuning-based}. A representative example is DiffusionCLIP~\citep{kim2022diffusionclip}, which leverages alignment in the CLIP semantic space to direct the denoising model. This method ensures that the displacement vectors of text embeddings before and after editing are matched by the displacement vectors of image embeddings, thereby enforcing semantic consistency in the edits. Although DiffusionCLIP targets specific concepts such as style transfer rather than individual images, it is classified as a test-time fine-tuning method due to its limited generalizability compared to training-time fine-tuning approaches.

Imagic~\citep{kawar2023imagic} also implements editing by aligning the semantic distance vectors of the text and corresponding image embeddings before and after editing. Unlike DiffusionCLIP, Imagic does not utilize pre-edit embeddings. Instead, it fixes the network parameters and initializes a post-edit embedding, \( e_{\text{tgt}} \), as a learnable parameter. The original image is optimized to correspond to this embedding, termed \( e_{\text{opt}} \). Subsequently, Imagic ensures that the semantic distance vectors of the pre-edit and post-edit images are similar. The fine-tuning process involves adjusting the entire denoising network to capture more details of the original image and the space where \( e_{\text{opt}} \) resides. Finally, the edited image is generated by inputting \( e_{\text{tgt}} \) into the fine-tuned model. Additionally, Imagic supports varying the editing intensity by inputting the difference between \( e_{\text{tgt}} \) and \( e_{\text{opt}} \).

\subsection{Adapter-Based Methods}
\label{sec:Adapter-Based Methods}

\begin{figure}[t]
\centering
\includegraphics[width=0.5\textwidth, trim=0cm 0.5cm 9cm 0cm, clip]{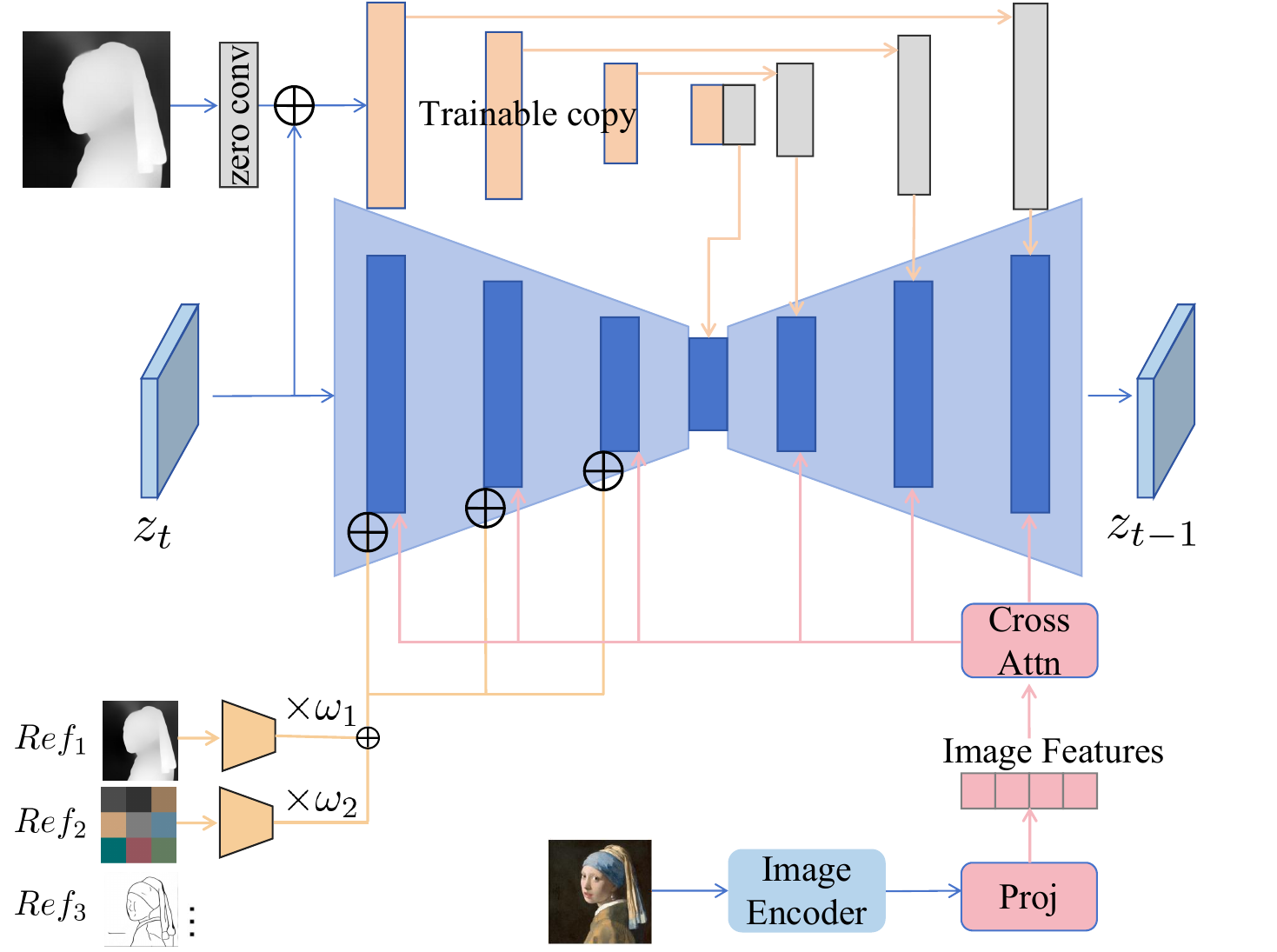} 
\caption{Three representative adapter structure: ControlNet~\citep{zhang2023adding}, IP-Adapter~\citep{ye2023ip} and T2I-Adapter~\citep{mou2024t2i}.}
\label{fig:adapter-based}
\end{figure}

As described in Sec.~\ref{sec:Tasks Definition}, editing instructions can be conveyed not only through text but also through images. 
In order for the diffusion model to better understand these multimodal prompts, a robust information processing module becomes essential
Adapters can efficiently process image information and integrate the processed data into the base model to facilitate finer editing. 
In the following, we first introduce several forms of adapters in controlled generation to establish a basic understanding of adapters. Then, we will describe how adapters can be used in image editing, and introduce the most important difference between them and controlled generation, i.e., the processing of the original image.

\subsubsection{Adapter for Controllable Generation}
Even though lots of the existing adapter-based methods are used to implement feature image to natural image conversion (controllable generation task in our definition), this way of processing image information is still useful for editing tasks. We have selected several representative methods based on the various approaches adapters employ to process images as shown in Fig.~\ref{fig:adapter-based}, and describe them in detail below.

The first category of adapters utilize a single structure to handle different categoric input images. A prominent example of this approach is ControlNet~\citep{zhang2023adding}. ControlNet operates by creating a trainable copy of the base model and training adapters with distinct weight values tailored to different classes of input images. During the processing of input images, ControlNet accesses the denoising model through a zero convolution (zero conv) method. The primary advantage of this technique is that the adapter component does not influence the output during the initial stages of training, thereby ensuring the stability of the generated images. Building upon this foundation, ControlNet++~\citep{li2025controlnet} introduces multiple enhancements and expansions to ControlNet, including a novel architecture and a consistency feedback mechanism, which significantly enhance the efficiency and consistency of conditional control.

Another class of adapters is dedicated to specific images. IP-Adapter~\citep{ye2023ip} is designed for processing portrait inputs. It incorporates a trainable projection module that processes the input image through a frozen image encoder to obtain image features. Unlike ControlNet, IP-Adapter utilizes cross-attention mechanisms to integrate the processed information into the base model, thereby retaining more detailed image features. Following this approach, AnyDoor~\citep{chen2024anydoor} further decomposes the input image into identifiers (IDs) and details, which are then fed into the base model separately, allowing for more granular control over the editing process.

In the aforementioned works, models typically handle a single input image. Furthermore, some approaches have advanced this capability by designing adapters that can process multiple images. Similar to ControlNet, T2I-Adapter~\citep{mou2024t2i} trains different adapters for various classes of conditions, but it weights these conditions for common control before feeding the processed image information into the base model. This method implements conditional weighting as follows:
\begin{equation}
h_{\text{fusion}} = \sum_{i=1}^n w_i \cdot \text{Adapter}_i(x_i), 
\end{equation}
where \(w_i\) denotes condition weights. Additionally, T2I-Adapter integrates the processed condition features into the intermediate features of the denoising network layer to facilitate the injection of control information.

UniControl~\citep{qin2023unicontrol} builds upon the trainable copy setup of ControlNet by introducing a Mixture-of-Experts (MoE) module, which dynamically selects different adapters based on the input conditions. Furthermore, Uni-ControlNet~\citep{zhao2024uni} proposes a method to merge adapters with different conditions by classifying them into local control adapters and global control adapters. This classification significantly reduces the number of adapters required. Moreover, processing similarly conditioned images with the same adapter enhances the training speed of the model by leveraging shared parameters.

\begin{figure}[t]
\centering
\includegraphics[width=0.5\textwidth, trim=0cm 0cm 12cm 1cm, clip]{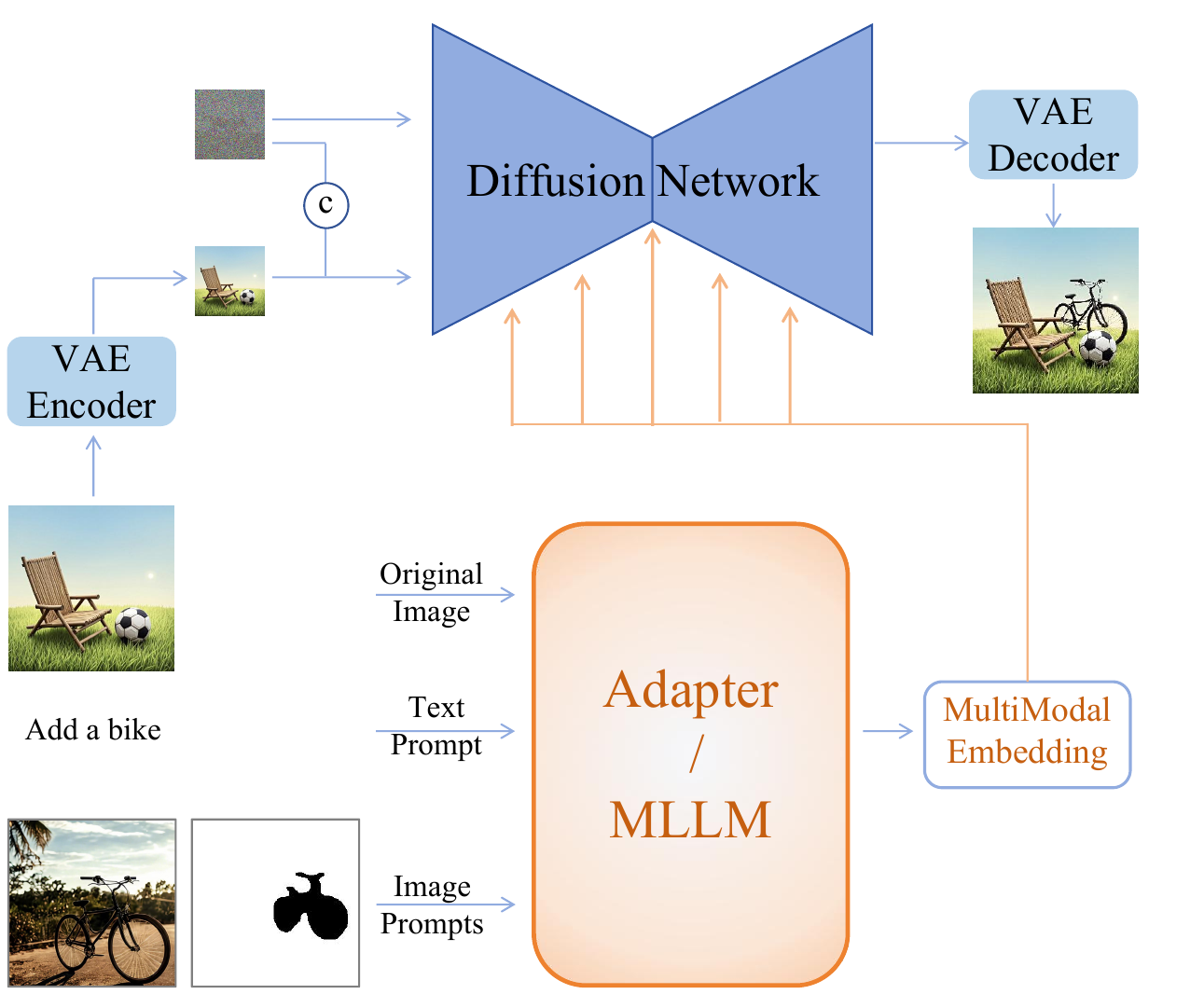} 
\caption{Original image and multimodal prompts processing based on adapter. Original image can be fed into the diffusion network by channel dimension concatenation~\citep{fu2023guiding, he2024freeedit} or token dimension concatenation with noise latent. All input could be processed using an Adapter/MLLM and then injected to diffusion network.}
\label{fig:adapter-based-edit}
\end{figure}

\subsubsection{Adapter for Image Editing}

Image editing and controlled generation differ significantly. Image editing tasks typically involve modifying the original image while retaining its details and the identity of its subjects. The inputs for image editing tasks are diverse and may include text instructions, natural reference images, and feature reference images (multimodal prompts). Thus, incorporating original image information and processing multimodal prompts are pivotal challenges. We illustrate how related work addresses these two aspects in Fig.~\ref{fig:adapter-based-edit}. Note that some recent methods simultaneously use adapters to process multimodal prompts and fine-tune the original model parameters. However, their main innovation and contribution lie in the processing of multimodal prompt information.

Various methods have been proposed to better preserve the details of the original image. MGIE~\citep{fu2023guiding} and FreeEdit~\citep{he2024freeedit} utilize the InstructPix2Pix strategy in Sec.~\ref{sec:Training Time Fine-tuning} for injecting original image information in image editing. OmniEdit~\citep{wei2024omniedit} integrates the MMDiT structure from SD3, which allows flexible text editing commands to more effectively align with the image representation space. This method retains additional image details by injecting noise latent representations as an auxiliary input. UniVG~\citep{ruan2024univg} concatenates original images with noise latent representations at the channel level and combines them with other multimodal prompts at the token level, supporting additional tasks by introducing conversion data from natural images to feature images.

Regarding the processing of multimodal prompts, BLIP-Diffusion~\citep{li2023blip} considers the original image as a type of prompt. It uses MLLM and q-former to extract the main subject of the original image as a subject prompt embedding, which is then input along with text prompt embeddings into the text encoder to enable the injection of multimodal prompt information. MGIE expands the architecture of BLIP-Diffusion, inputting editing instructions into MLLM and employing a learnable adapter to process images, thereby flexibly selecting visual tokens and injecting them via cross attention. FreeEdit uses U-Net to process the reference image into multi-level features, injected layer by layer. DreamOmni~\citep{xia2024dreamomni} uses VLM to uniformly process the original image, image prompts, and text prompts. It combines original images processed through VAE and noise-added latent representations, then concatenates these with VLM features at the token level within the DiT module. MIGE~\citep{tian2025mige} processes the original image and image prompts using CLIP and VAE to obtain semantic and visual features, respectively. The visual tokens obtained are processed alongside editing instructions by an LLM and injected into the diffusion network using cross-attention, thereby supporting more reference images.

\section{Evaluation}
\label{sec:Evaluation}

\begin{figure}[t]
\centering
\includegraphics[width=1\textwidth, trim=0cm 5.5cm 7cm 0cm, clip]{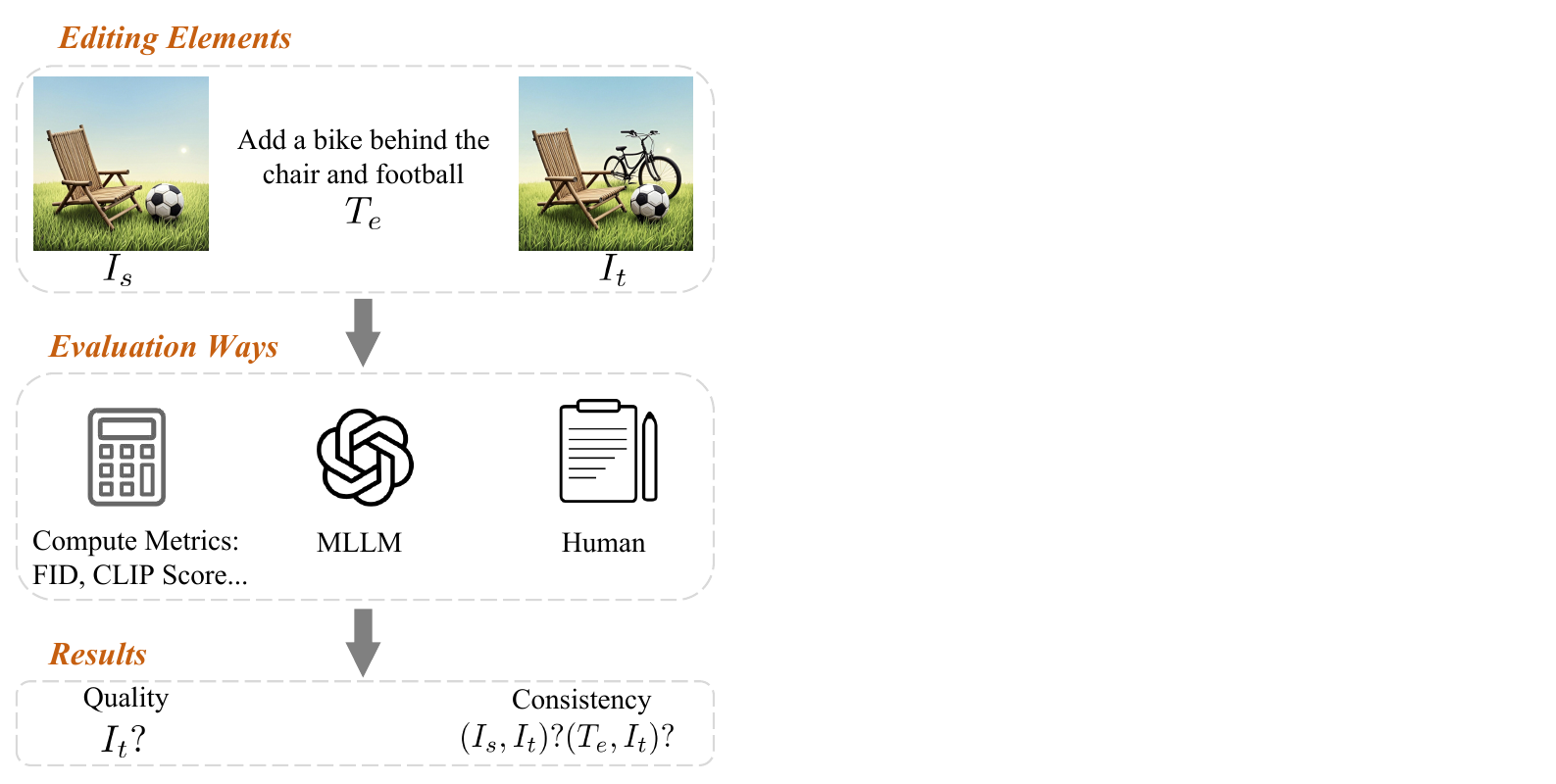} 
\caption{Performance assessment pipeline. Evaluation ways encompasses three approaches: computational metrics, Multimodal Large Language Models (MLLM), and human evaluations. $I_s$, $I_t$, and $T_e$ represent the source image, the edited image and the text editing instruction.}
\label{fig:dataset construction}
\end{figure}

\subsection{Metrics}
\label{sec:Metrics}

To comprehensively evaluate the performance of image editing models, it is essential to assess their editing effects. We presents the performance assessment pipeline in Fig.~\ref{fig:dataset construction}. Methods for obtaining evaluation metrics can be categorized into three types: firstly, compute metrics, which involve scoring by calculating indicators such as FID~\citep{heusel2017gans} and CLIP Score~\citep{hessel2021clipscore}. This approach is relatively objective and follows standardized computational procedures, but due to the flexibility inherent in image editing, the scores may lack precision; secondly, utilizing Multimodal Large Language Models (MLLM) to directly score the editing results, where the key focus lies in the design of the scoring mechanism; thirdly, employing human evaluation. A common approach in human evaluation is to provide standardized inputs to different models and then compare the output images generated by each model~\citep{hertz2022prompt}. In this process, researchers typically recruit a sufficient number of evaluators to score both the input images and those generated by different models, ensuring that evaluators remain blind to the origin of each image to guarantee the fairness of the evaluation. Finally, the scores from all evaluators are aggregated to compute the final scores for each model across various evaluation dimensions. After evaluation ways, we will introduce two important evaluation dimensions and their related metrics below.

\subsubsection{Quality}
The evaluation of image editing models can be categorized into two primary aspects: quality and consistency. To assess image quality, the Fréchet Inception Distance (FID)~\citep{heusel2017gans} is a widely used metric. FID measures the similarity between the distribution of generated images and that of real images. Specifically, this metric involves extracting features from images using a pre-trained model like InceptionV3~\citep{szegedy2015going} and computing the distance between the mean and covariance matrices of the feature distributions of generated and real images. A lower FID value indicates higher quality of generated images and a closer alignment with real images.

Other similar metrics include Kernel Inception Distance (KID)~\citep{binkowski2018demystifying} and Learned Perceptual Image Patch Similarity (LPIPS)~\citep{zhang2018unreasonable}. KID is analogous to FID but employs a kernel-based approach to compute Maximum Mean Discrepancy (MMD), offering a more stable alternative for image quality assessment. LPIPS is a perceptual metric that evaluates image similarity by comparing feature representations from different layers of a pre-trained VGG network, closely aligning with human visual perception.

Additionally, researchers have introduced other quality assessment metrics, such as the DINO~\citep{caron2021emerging} metric. This metric evaluates visual consistency by calculating the average cosine similarity between the DINO embeddings of generated images and those of real images using the Vision Transformer (ViT-S/16) model~\citep{ruiz2023dreambooth}. This provides further insight into the visual quality of the generated images.

\subsubsection{Consistency}

For consistency evaluation, two primary dimensions are considered: consistency between the edited and original images, and consistency between the output images and the editing instructions. CLIP score~\citep{hessel2021clipscore} is a commonly employed metric for assessing such consistency. The CLIP model~\citep{radford2021learning} effectively aligns the latent spaces of text and images, where the similarity or distance between text and images in this aligned space reflects the degree of alignment between generated images and textual descriptions.

Specifically, P2P~\citep{hertz2022prompt} utilizes CLIP Text-image direction similarity to measure the alignment between generated images and textual descriptions. In contrast, DreamBooth~\citep{ruiz2023dreambooth} and IP-Adapter~\citep{ye2023ip} employ CLIP-I and CLIP-T, respectively, to further refine the evaluation of consistency between generated images and textual instructions. These approaches leverage the CLIP model with distinct computational methods to assess the degree of alignment between generated images and the provided editing instructions.

Furthermore, to evaluate the visual consistency between feature images and natural images, Uni-ControlNet~\citep{zhao2024uni} incorporates multiple evaluation metrics tailored to different conditions. For tasks involving edge detection (Canny, HED, MLSD) and sketch generation, the Structural Similarity Index (SSIM)~\citep{wang2003multiscale} is used to assess image quality by considering brightness, contrast, and structural features, thereby providing a measure that closely aligns with human visual perception. To evaluate pose estimation accuracy, Uni-ControlNet employs mean Average Precision (mAP) based on Object Keypoint Similarity (OKS). For depth estimation tasks, Mean Squared Error (MSE) is utilized to measure the discrepancy between generated and real depth maps. Lastly, Mean Intersection over Union (mIoU)~\citep{rezatofighi2019generalized} is applied to assess the quality of segmentation maps, reflecting the overlap between output images and real segmentation masks. These metrics collectively ensure a comprehensive evaluation of image editing tasks.

\begin{figure*}[t]
\centering
\includegraphics[width=1\textwidth, trim=0cm 11cm 4cm 0cm, clip]{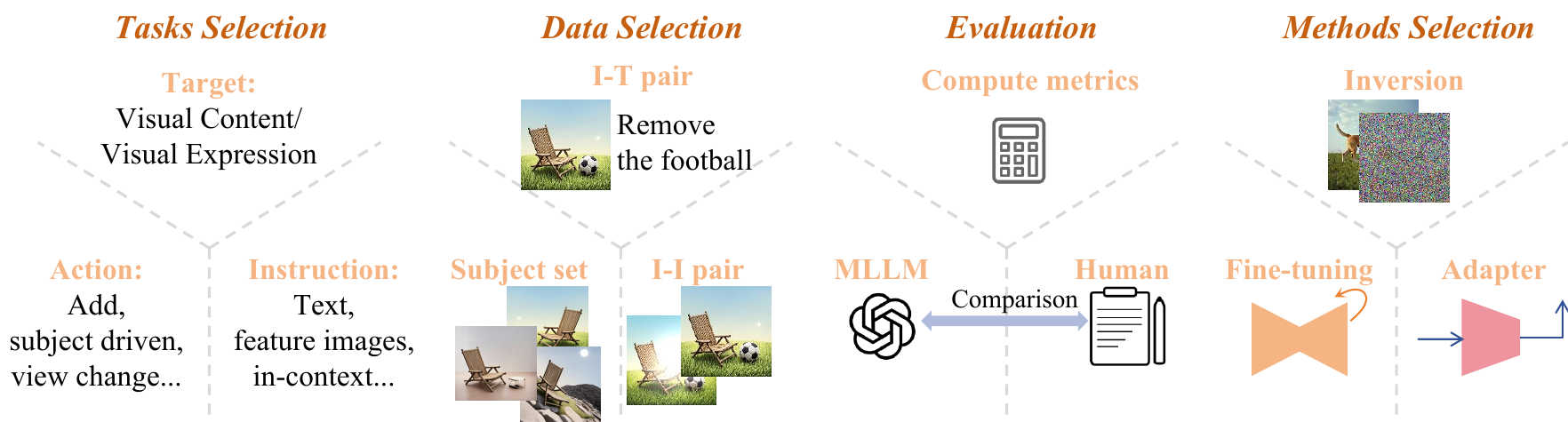} 
\caption{General pipeline of editing benchmarks.}
\label{fig:Benchmark pipeline}
\end{figure*}

\subsection{Benchmarks}
\label{sec:Benchmarks}

In the evaluation of image editing models, metrics serve as crucial tools for quantifying performance and assessing strengths and weaknesses. However, due to the diversity of evaluation approaches, establishing a unified assessment system is essential for reliable comparisons and research reproducibility. To this end, several specialized image editing benchmarks have been developed, each offering standardized testing frameworks and evaluation protocols that enhance model comparison and provide valuable references for research and practice. 

EditBench~\citep{wang2023imagen} is a systematic benchmark targeting text-guided inpainting tasks, evaluating objects, attributes, and scenes to ensure comprehensive testing. It has demonstrated the effectiveness of object masking in enhancing consistency between text prompts and generated images.
EditVal~\citep{basu2023editval} serves as a standardized benchmark for text-guided image editing, covering 13 edit types and employing pre-trained vision-language models for automated evaluation, with results closely aligning with human preferences.
TEdBench~\citep{kawar2023imagic} evaluates complex text-guided editing methods, showcasing the ability to perform intricate object edits while preserving image characteristics.
I2EBench~\citep{ma2024i2ebench} is a comprehensive framework spanning 16 evaluation dimensions, incorporating user studies to align results with human perception.
EditEval~\citep{huang2024diffusion}, tailored for diffusion models, contains 50 high-quality images and prompts for seven common tasks, utilizing LMM Score and user studies for thorough assessment. 

Despite their distinct focuses, these benchmarks follow a general evaluation pipeline as summarized in Fig.~\ref{fig:Benchmark pipeline}, including tasks selection, data selection, evaluation and methods selection. In the subsequent sections, we introduce these benchmarks following this order in detail.

\subsubsection{Tasks and Data Selection}

Given the diversity of editing tasks, it is crucial to evaluate the effectiveness of image editing models across multiple task types. While Imagic~\citep{kawar2023imagic} claims to achieve complex non-rigid edits, there was no established standard benchmark for evaluating non-rigid text-based image editing at the time. Consequently, TEdBench was introduced, consisting of 100 pairs of input images and target texts describing desired complex non-rigid edits. This underscores the necessity of testing across various editing task types. However, the scope of TEdBench remains limited in comparison to I2EBench, which categorizes editing tasks into High-level Editing and Low-level Editing. High-level editing includes tasks such as Object Removal, Background Replacement, Color Alteration, and Style Alteration, while Low-level editing focuses on tasks like Deblurring, Rain Removal, Haze Removal, and others.

Observations from EditEval reveal that most methods perform well on semantic and stylistic tasks but face challenges with structural editing due to difficulties in achieving accurate spatial awareness in current text-to-image (T2I) diffusion models. To precisely assess and differentiate the capabilities of various editing models, EditEval selected seven common tasks: Object Addition, Object Replacement, Object Removal, Background Replacement, Style Change, Texture Change, and Action Change. These tasks aim to comprehensively evaluate the performance of editing methods, ranging from simple object edits to complex scene modifications.
OmniEdit~\citep{wei2024omniedit} selects similar editing tasks, including seven different skills like addition, swapping, removal, attribute modification, background change, environment change, and style transfer.

The selection of test data is an integral part of benchmark design and is closely tied to the types of editing tasks. TEdBench constructed a novel dataset of 100 input-image and target-text pairs to evaluate complex non-rigid edits. In contrast, I2EBench meticulously curated approximately 140 images from publicly available datasets~\citep{lin2014microsoft, guo2023sky, liu2021synthetic, nah2017deep} for each task, ensuring a wide diversity of subjects and scenes. These images were annotated with textual editing instructions by human annotators, and to further enhance instruction diversity, I2EBench employed ChatGPT~\citep{achiam2023gpt} to rewrite the original instructions. EditEval manually selected 50 images from Unsplash’s online repository of professional photographs, ensuring a broad range of themes and scenarios. These images were categorized into seven groups, each corresponding to one of the specific editing tasks mentioned above. A detailed template was designed, and GPT-4V was utilized to generate initial editing prompts and instructions, which were subsequently reviewed to ensure clarity, specificity, and applicability.
Reason-Edit~\citep{huang2024smartedit} argues that a good instruction-based editing model requires some reasoning ability to better understand instruction intent. Therefore, it selects instructions that require different levels of reasonina, such as describing the object to be edited through positional relationships, interaction relationships, attributes, or functions. Editing models need to be locked to the editing object first according to these descriptions in order to achieve precise editing effects.

By carefully selecting and categorizing tasks and data, these benchmarks provide a systematic and standardized approach to evaluating image editing models, thereby facilitating advancements in research and practice.

\subsubsection{Evaluation and Methods Selection}

Once the evaluation tasks and datasets have been established, the next step involves assessing the output quality of each model. TEdBench employs a strategic evaluation approach where participants are presented with an input image and a target text, then asked to select the better editing result from two options using the Two-Alternative Forced Choice (2AFC) method~\citep{kolkin2019style, park2020swapping}. While effective, this approach demands significant resources and becomes impractical for large-scale testing due to its reliance on human participants. 

With advancements in Multimodal Large Language Models (MLLMs)~\citep{liu2024sphinx, chu2024mobilevlm, ma2022x}, such as GPT-4V~\citep{achiam2023gpt}, Gemini Pro~\citep{team2024gemini}, and QWen-VL~\citep{bai2023qwen}, which exhibit enhanced capabilities in automated image understanding, I2EBench leverages these models to score the majority of high-level editing tasks. Additionally, I2EBench incorporates human evaluations to validate the effectiveness of automated model evaluations by demonstrating a strong correlation between human assessments and those provided by large language models. This dual approach ensures comprehensive and reliable assessments.

EditEval further refines the use of large language models for scoring by designating four distinct evaluation dimensions: editing accuracy, contextual preservation, visual quality, and logistic realism. Scores are obtained for each dimension and then weighted according to specific proportions to derive a comprehensive final score. Similar to I2EBench, EditEval employs human evaluations to validate the effectiveness of their evaluation framework, offering a more detailed examination of the results.

In terms of methodology selection, TEdBench primarily compares three methods: SDEdit~\citep{meng2021sdedit}, DDIB~\citep{su2022dual}, and Text2Live~\citep{bar2022text2live}. In contrast, I2EBench expands its comparative analysis by including additional methods such as InstructAny2Pix~\citep{li2023instructany2pix}, HIVE~\citep{zhang2024hive}, InstructEdit~\citep{wang2023instructedit}, InstructDiffusion~\citep{geng2024instructdiffusion}, InstructPix2Pix~\citep{brooks2023instructpix2pix}, MagicBrush~\citep{zhang2024magicbrush}, MGIE~\citep{fu2023guiding}, and HQEdit~\citep{hui2024hq}. EditEval emphasizes methodological diversity by selecting approaches based on different training procedures, including training-based methods, methods requiring fine-tuning, and those that are training and fine-tuning free. 

The selection criteria ensure fairness and consistency: the methods must rely solely on text conditions, possess the capability to handle specific tasks, and provide accessible open-source code. To avoid limitations stemming from restricted applicability across domains, domain-specific methods are excluded. The final selections include models such as InstructPix2Pix, Instruct-Diffusion, Imagic~\citep{kawar2023imagic}, DDPM-Inversion~\citep{huberman2024edit}, LEDITS++~\citep{brack2024ledits++}, ProxEdit~\citep{han2024proxedit}, and AIDI~\citep{pan2023effective}. This diverse range of methods ensures a thorough and representative evaluation benchmark.

\section{Datasets}
\label{sec:Datasets}

In the rapidly development of computer vision and image editing, the availability and quality of training datasets play a pivotal role in shaping the performance and capabilities of models. Existing editing methods have leveraged a diverse range of data sources, each with its unique characteristics and construction processes. However, the complexity and variability of real-world editing tasks necessitate the development of more advanced and targeted datasets. This section aims to provide a comprehensive overview and classification of the existing approaches to acquiring and constructing training datasets for image editing. By understanding these methods, researchers can better utilize available data and design more effective models tailored to specific editing requirements.

\begin{figure*}[t]
\centering
\includegraphics[width=1\textwidth, trim=0cm 7cm 0cm 0cm, clip]{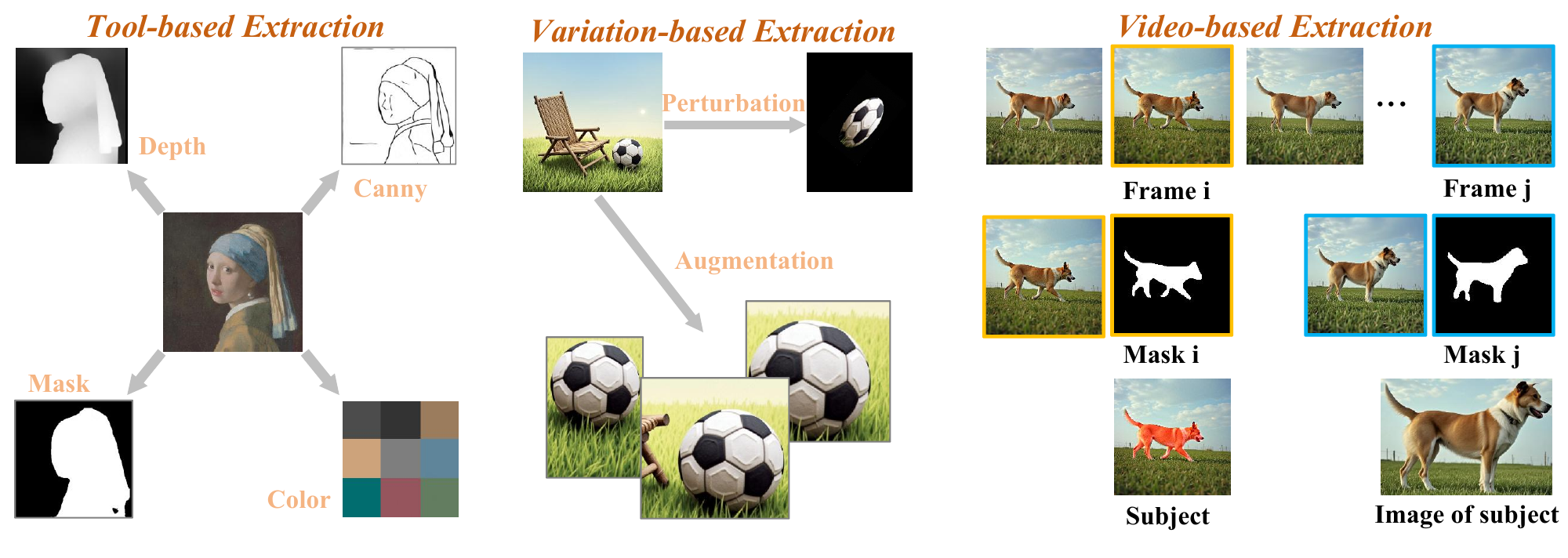} 
\caption{Three types of extraction-based dataset construction methods.}
\label{fig:extraction-based dataset construction}
\end{figure*}

\subsection{Extraction-based Dataset Construction}

The field of computer vision has amassed an extensive collection of image datasets, which serve as the foundation for numerous applications. However, these datasets often fall short when it comes to the specific needs of image editing tasks. Unlike traditional vision tasks, image editing requires paired images that capture the same scene or object before and after an edit, presenting a distinct challenge. This has led to the development of specialized methods for constructing image pairs that can effectively support editing tasks. In the following, we will introduce the approaches used by different methods to extract and process existing datasets as shown in Fig.~\ref{fig:extraction-based dataset construction}, which creates valuable resources for image editing research.

\subsubsection{Feature Image Extraction}

To construct feature image and natural image pairs, ControlNet~\citep{zhang2023adding} integrates various feature extraction models and datasets to get a comprehensive collection of feature image pairs. For example, for the canny edge feature, ControlNet collects approximately three million images from the internet and applies the canny edge detector~\citep{canny1986computational} to extract edge maps, thereby forming edge-image description pairs. These pairs are then sorted by resolution and divided into subsets to facilitate further testing of the impact of data scale. Additionally, ControlNet employs a deep learning-based Hough transform~\citep{gu2022towards} on the Places2~\citep{zhou2017places} dataset to detect straight lines and integrates the BLIP~\citep{li2022blip} model to generate image descriptions. Due to the similarity between Hough lines and canny edges, the pre-trained canny edge adapter can be reused for learning Hough line features. For other features, ControlNet adopts similar methodologies. For instance, it utilizes HED boundary detection~\citep{xie2015holistically} to generate HED boundary maps, the MiDaS~\citep{lasinger2019towards} model to produce depth maps, and learning-based pose estimation methods~\citep{cao2017realtime, kreiss2021openpifpaf} to extract human pose maps. Through data augmentation~\citep{vinker2023clipascene}, HED boundary maps are further transformed into user sketches. For semantic segmentation and normal maps, ControlNet directly leverages existing datasets such as COCO-Stuff~\citep{caesar2018coco}, ADE20K~\citep{zhou2017scene}, and DIODE~\citep{vasiljevic2019diode}, with captions generated by the BLIP model.

The data construction process of T2I-Adapter~\citep{mou2024t2i} is relatively simpler and more unified. When constructing sketch maps, it uses the COCO14~\citep{lin2014microsoft} dataset, which contains 164K images, as the foundation and employs a predefined edge prediction model~\citep{su2021pixel} to generate framework maps. For semantic segmentation maps, it directly utilizes the COCO-Stuff dataset, which includes 80 object classes and 91 scene classes. For keypoints, color, and depth maps, T2I-Adapter uniformly selects 600,000 image-text pairs from the LAION AESTHETICS dataset, integrating the MMPose~\citep{contributors2020openmmlab} model to extract keypoint maps and the MiDaS~\citep{ranftl2020towards} model to generate depth maps. Furthermore, color block maps are created using blurring techniques.

Since ControlNet’s training dataset has not been publicly released, UniControl~\citep{zhao2024uni} has built a large-scale dataset from scratch, named MultiGen-20M. This dataset is based on a subset of Laion-Aesthetics-V2~\citep{schuhmann2022laion}, filtering images with aesthetic scores above 6 and resolutions of no less than 512 pixels, resulting in approximately 2.8 million image-text pairs. Subsequently, UniControl processes this dataset across multiple dimensions, covering five categories (edges, regions, skeletons, geometric maps, real images) and nine specific tasks. For example, for Canny, HED, Depth, and Normal features, it employs the Canny edge detector, Holistically-Nested Edge Detection~\citep{xie2015holistically}, MiDaS, and OpenPose~\citep{cao2017realtime}, respectively. For segmentation maps and object bounding boxes, it uses the Uniformer model pre-trained on ADE20K and the YOLOv4~\citep{bochkovskiy2020yolov4} model trained on COCO. Additionally, UniControl creates boundary masks for source images with random masking percentages ranging from 20\% to 80\% for image outpainting tasks.

\subsubsection{Subject-Driven Data Extraction}

Data pairs or sets of same objects in different scenarios constitute a significant category of editing datasets, which are crucial for subject-driven tasks. Earlier methods, such as DreamBooth~\citep{ruiz2023dreambooth}, did not require training time fine-tuning with the need for extensive training data. However, to support training time fine-tuning paradigms, larger and more comprehensive datasets are necessary. Below, we outline the construction processes for such datasets.

ObjectStitch~\citep{song2023objectstitch} collects synthetic training data from Pixabay and utilizes an object instance segmentation model~\citep{lee2020centermask} to predict panoptic segmentation and classification labels. The dataset is first filtered to remove objects that are either too small or too large. Inspired by \citet{detone2016deep}, the four points of each object’s bounding box are randomly perturbed to apply projective transformations, followed by random rotations within the range of $[-20^\circ, 20^\circ]$ and color perturbations. The segmentation mask is then used to extract the object. One advantage of this approach is the elimination of manual annotation, as the original image serves as the ground truth. Additionally, using the bounding box as a mask not only fully covers the object but also includes its neighboring regions, allowing space for generating shadows. This method proves sufficiently flexible, enabling the model to apply spatial transformations, synthesize new views, and generate shadows and reflections. Furthermore, Paint by Example~\citep{yang2023paint} incorporates this data augmentation method into its training process.

Previous studies~\citep{song2023objectstitch, yang2023paint} typically utilized single images and applied data augmentation techniques such as rotation, flipping, and elastic transformations. However, these simplistic methods struggle to adequately represent the diverse poses and viewpoints of objects in real-world scenarios. To address this limitation, AnyDoor~\citep{chen2024anydoor} uses video datasets by selecting frames of the same object in different scenes. Specifically, video segmentation and tracking technologies are used to extract segments where a specific object appears. Two frames are selected, and the foreground object’s mask is extracted. Subsequently, the background of one frame is removed, and the object is cropped based on the mask to obtain the target object image. This mask, after perturbation, is used as a control mask. For the other frame, a bounding box is generated, and the area within the box is removed to obtain the scene image, while the original image without the mask serves as the ground truth for training. AnyDoor’s video data encompasses a wide range of domains, including natural scenes, virtual try-on, and multi-view objects~\citep{wang2021unidentified, miao2022large, yu2023mvimgnet, choi2021viton}. 
FramePainter~\citep{zhang2025framepainter} approaches image editing by treating it as a video generation task. It considers two frames from a video as an image pair and applies various methods to extract editing instructions. These methods include SEA-RAFT~\citep{wang2024sea} for optical flow analysis, CoTracker-v3~\citep{karaev2024cotracker3} for point tracking, and the Sobel filter~\citep{zhan2019self} for sketch extraction, among others.

By employing advanced data construction techniques and leveraging diverse datasets, these methods ensure the creation of robust and varied data pairs or sets. This diversity is essential for training image editing models to handle a wide range of real-world scenarios, thereby enhancing their performance and generalization capabilities in subject-driven tasks.

\begin{figure*}[t]
\centering
\includegraphics[width=1\textwidth, trim=0cm 0cm 2cm 0cm, clip]{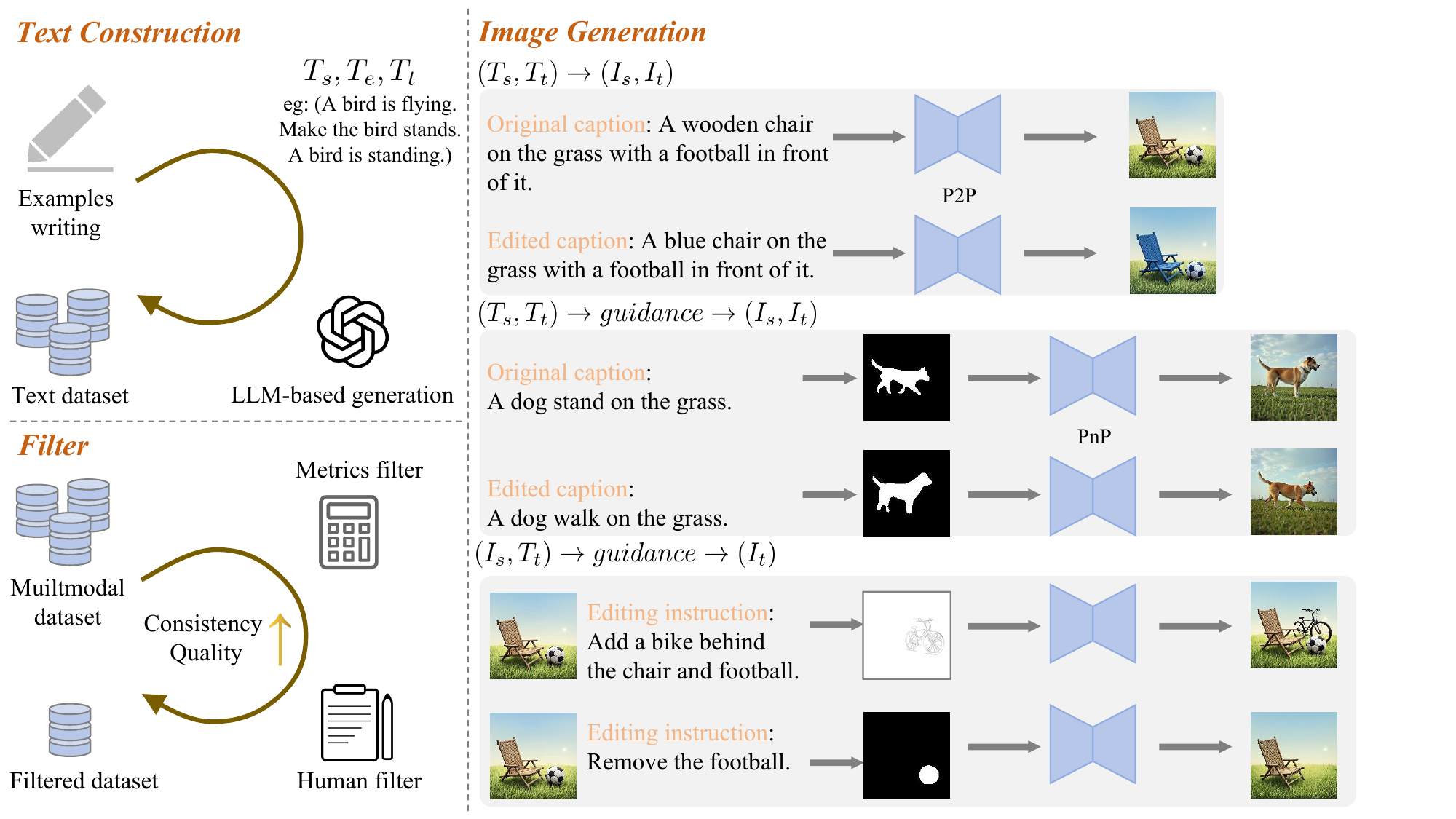} 
\caption{Generation-based dataset construction pipeline. The ``guidance" in image generation part includes mask, canny and layout, etc. The P2P and PnP refer to \citet{hertz2022prompt} and \citet{tumanyan2023plug}.}
\label{fig:generation-based dataset construction}
\end{figure*}

\subsection{Generation-based Dataset Construction}

While extraction-based methods have made significant strides in constructing datasets for image editing, they often lack the flexibility and richness of instructions required for more complex editing tasks. To address this limitation, generation-based dataset construction approaches have emerged, aiming to create datasets with more diverse and detailed textual instructions. These methods leverage the power of large language models and text-to-image generation techniques to produce datasets that can guide models in performing a wider range of editing operations. In this subsection, we will explore the key components of generation-based dataset construction, including text instruction construction, image pair generation, and data filtering, as shown in Fig.~\ref{fig:generation-based dataset construction}. We will present how different methods, such as InstructPix2Pix~\citep{brooks2023instructpix2pix}, Emu Edit~\citep{sheynin2024emu}, and OminiControl~\citep{tan2024ominicontrol}, perform these steps, and the strategies they employ to ensure the quality and consistency of the generated datasets.

have approached these steps and the strategies they employ to ensure the quality and consistency of the generated datasets.

\subsubsection{Text Instruction Construction}
The first step in constructing an editing dataset is to generate pre-edit and post-edit texts along with corresponding textual editing instructions. Following the introduction of P2P~\citep{hertz2022prompt}, which facilitates editing without requiring training, InstructPix2Pix~\citep{brooks2023instructpix2pix} utilized this approach to build an editing dataset. Initially, 700 input captions were extracted from the LAION-Aesthetics V2 6.5+ dataset~\citep{schuhmann2022laion}, and instructions along with output captions were manually crafted. This approach established a pattern for image editing by using triplets of (input caption, editing instruction, output caption) to fine-tune GPT-3. Leveraging GPT-3's extensive knowledge and generalization capabilities, the fine-tuned model was able to generate creative and coherent instructions and captions.

Emu-edit~\citep{sheynin2024emu} differs from InstructPix2Pix in its construction of text instructions by replacing GPT-3 with Llama 2. Additionally, it was observed that using a single agent to generate instructions for all tasks resulted in a lack of diversity in the dataset. Large language models, in particular, tend to exhibit biases toward specific tasks and phrasing of instructions. To address this issue, emu-edit separated different editing tasks to ensure each was adequately represented with sufficient data. Furthermore, they enhanced the diversity of the text instructions generated by the large model by performing three operations on the example text triplets: (1) shuffling among examples, (2) randomly sampling 60\% of the examples, and (3) randomly changing the verbs in the examples from a predefined set of words.

OminiControl~\citep{tan2024ominicontrol} also employed a generative approach to construct an editing dataset named Subject200K, specifically targeting subject-driven tasks. Unlike previous approaches, this dataset was created using GPT-4~\citep{achiam2023gpt} to generate textual descriptions. They selected 42 diverse object categories, including furniture, vehicles, electronics, clothing, and others, and created multiple object instances for each category, totaling 4,696 unique objects. Each object entry consists of: (1) a brief description, (2) eight diverse scene descriptions, and (3) one studio photo description. Notably, while the dataset for subject-driven tasks mentioned in the previous section was generated through extraction, Subject200K was constructed through generation. Although it targets different editing tasks, the underlying construction process remains similar.

By utilizing advanced generative models and incorporating diversity-enhancing techniques, these approaches ensure the creation of robust and varied datasets for training image editing models. This variety is essential for developing models capable of handling a wide range of editing tasks with high accuracy and coherence.

\subsubsection{Image Generation}

After obtaining the textual descriptions, the next step is to generate corresponding image pairs based on these descriptions. However, when directly using text-to-image models, even slight differences in the prompt text can lead to inconsistencies in the generated images. This inconsistency is unsuitable for constructing an editing dataset, as the goal is to use these paired data as supervision for training models to edit images (rather than generating different random images). Therefore, InstructPix2Pix selected P2P to generate images. When the textual differences are minor, P2P can maintain a certain level of consistency in non-edited target areas. Additionally, P2P offers a mixture ratio parameter that controls the intensity of the edits. By setting a lower edit degree, it ensures a degree of consistency even when the textual differences are significant.

However, in most image editing tasks, word-to-word alignment is not a practical assumption. This approach often fails to preserve structure and identity, and reducing the editing intensity can limit the effectiveness of the edits. To address this challenge, emu-edit proposed using masks to precisely control the editing regions, thereby constructing better data for region-specific editing tasks. It first compares the differences between the pre-edit and post-edit texts to identify the objects to be edited, then uses DINO to obtain the mask. During the generation process of P2P, it merges two noise latents according to the mask to create the latent corresponding to the post-edit text, ensuring precise reconstruction of the masked area. In practice, emu-edit found that this hard thresholding could cause some edited image generations to fail. Consequently, it proposed using three types of masks: (1) precise masks obtained through DINO and SAM~\citep{kirillov2023segment}, (2) gaussian-blurred expansion of the first type of mask's boundaries, and (3) bounding box masks. All images generated using these three methods were retained for the next filtering step. Building on Emu Edit, AnyEdit~\citep{yu2024anyedit} expanded control from masks to relevance maps and layouts. Additionally, it used a richer set of editing models, such as AnyDoor and SD-Inpaint~\citep{rombach2022high}, to construct the data. Formally, this approach differs from previous methods by using the original image and editing instructions to obtain the edited data.

Besides using editing models to obtain image pairs, direct use of text-to-image models is also possible. Since subject-driven tasks only require consistency in the main subject, OminiControl~\cite{tan2024ominicontrol} designed a specific prompt and selected FLUX as the text-to-image model. Specifically, the prompt input to FLUX is composed of three parts: ``Two side-by-side images of the same object: {brief\_description}", ``Left: {scene\_description1}", and ``Right: {scene\_description2}". This method significantly increases the consistency of the main subject in both images, allowing subsequent filtering to obtain the target data.

\subsubsection{Data Filtering}

After obtaining a large volume of generation-based image editing data, the final step is to select high-quality data from this pool. Generally, each dataset is scored across multiple dimensions, and data with low scores are filtered out. The scoring dimensions include image-to-image similarity, adherence to instructions, and others, as detailed in Sec.~\ref{sec:Evaluation}. Additionally, since the images are generated, some filtering processes used in constructing large image datasets are also employed. For example, in Emu, a series of automated filters are used to narrow down the pool to several hundred million images. These filters include, but are not limited to, removing offensive content, applying aesthetic score filters, and using optical character recognition (OCR) to filter images with excessive overlapping text.

Apart from directly using hard thresholds to filter out data, another approach is to assign different optimization weights based on the quality of the data. OMniEdit-Filtered-1.2M~\citep{wei2024omniedit} utilizes InternVL2~\citep{chen2024internvl} to distill the scoring capabilities of GPT-4, and then rates all the data. During training, the loss obtained from different data within a batch is weighted according to the data's scores. This approach can be regarded as a form of soft threshold data filtering.

\section{Challenges and Future Directions}
\label{sec:Challenges and Future Directions}

Recent research has made significant progress in the field of image editing. However, due to the diversity and complexity of editing tasks, existing methods still struggle to achieve the desired editing effects in certain scenarios. In this section, we discuss several unresolved issues and challenges in current approaches and propose potential future directions for researchers.

Inversion-based methods, while capable of achieving objective editing effects without extensive training, still face numerous challenges. Firstly, the inherent randomness in the diffusion process makes precise inverse operations difficult, often leading to a loss of detail. For instance, methods like SDEdit~\citep{meng2021sdedit} and Null-text~\citep{mokady2023null} inversion require carefully designed noise scheduling and loss functions, which may not be suitable for all types of editing tasks.

Fine-tuning-based methods, although versatile in handling a variety of editing tasks, require substantial annotated data and significant computational resources. The quality and diversity of datasets are crucial for training robust models, with the upper limit of editing performance being constrained by the available training data. Additionally, existing datasets, despite being rich in editing tasks, often lack diverse multimodal editing instruction data, making it challenging for trained models to handle more complex editing scenarios effectively.

Adapter-based methods introduce additional modules to process editing commands without modifying the base model's parameters. Despite their high flexibility, these methods encounter several challenges. For example, ControlNet and IP-Adapter require meticulously designed and trained adapter components to ensure optimal performance. Furthermore, current adapter methods may not efficiently handle multi-image inputs or complex editing tasks, as they typically focus on single-image editing.

Improved evaluation metrics are also crucial for advancing the field. Current metrics, such as FID and LPIPS, primarily focus on image quality and may not fully capture the nuances of editing tasks. This limitation can result in models performing well on these metrics while failing to produce semantically coherent editing results in practical applications. Moreover, existing metrics may not be suitable for evaluating complex editing tasks involving multiple images or conditions, such as video editing or multi-target operations.

Future editing models should support a wide range of editing tasks, eliminating the need to switch between multiple models for different tasks. Additionally, while current editing models primarily rely on text instructions, future models should support more multimodal input instructions for more precise editing. Given the commonalities among many tasks, which can mutually enhance each other, a promising trend is to integrate various editing and visual tasks~\citep{xiao2024omnigen}. Another emerging trend is enabling multi-turn image editing~\citep{lee2024semanticdraw}, where ensuring consistency across multiple editing rounds poses a significant challenge.

Addressing these challenges will require innovative approaches in model architecture, data acquisition, and evaluation methodologies. By focusing on these areas, future research can develop more versatile, efficient, and effective image editing models capable of handling the increasingly complex demands of real-world applications.

\section{Conclusion}
\label{sec:Conclusion}

This survey provides a comprehensive review of image editing techniques based on diffusion models, systematically organizing the current research progress across multiple dimensions, including task definition, method classification, evaluation metrics, and training datasets. Through in-depth analysis, this survey offers researchers a clear developmental trajectory of image editing technologies, revealing key techniques and research trends in the field.
Firstly, we clearly define image editing tasks and categorize various forms of editing tasks in detail from the perspectives of the object of operation and the method of operation. In terms of method classification, we divide existing editing methods into three major categories: inversion-based methods, fine-tuning-based methods, and adapter-based methods, providing detailed introductions and analyses of each. Additionally, we organize and summarize the commonly used evaluation metrics, available datasets, and their construction methods. Based on a review of existing research, we also offer insights into the future directions of development in the field of image editing.


\end{sloppypar}

\bibliographystyle{spbasic}
\bibliography{reference.bib}

\begin{thebibliography}{176}
\providecommand{\natexlab}[1]{#1}
\providecommand{\url}[1]{{#1}}
\providecommand{\urlprefix}{URL }
\expandafter\ifx\csname urlstyle\endcsname\relax
  \providecommand{\doi}[1]{DOI~\discretionary{}{}{}#1}\else
  \providecommand{\doi}{DOI~\discretionary{}{}{}\begingroup \urlstyle{rm}\Url}\fi
\providecommand{\eprint}[2][]{\url{#2}}

\bibitem[{Achiam et~al.(2023)Achiam, Adler, Agarwal, Ahmad, Akkaya, Aleman, Almeida, Altenschmidt, Altman, Anadkat et~al.}]{achiam2023gpt}
Achiam J, Adler S, Agarwal S, Ahmad L, Akkaya I, Aleman FL, Almeida D, Altenschmidt J, Altman S, Anadkat S, et~al. (2023) Gpt-4 technical report. arXiv preprint arXiv:230308774

\bibitem[{Avrahami et~al.(2022)Avrahami, Lischinski, and Fried}]{avrahami2022blended}
Avrahami O, Lischinski D, Fried O (2022) Blended diffusion for text-driven editing of natural images. In: Proceedings of the IEEE/CVF conference on computer vision and pattern recognition, pp 18208--18218

\bibitem[{Avrahami et~al.(2023)Avrahami, Fried, and Lischinski}]{avrahami2023blended}
Avrahami O, Fried O, Lischinski D (2023) Blended latent diffusion. ACM transactions on graphics (TOG) 42(4):1--11

\bibitem[{Bai et~al.(2023)Bai, Bai, Yang, Wang, Tan, Wang, Lin, Zhou, and Zhou}]{bai2023qwen}
Bai J, Bai S, Yang S, Wang S, Tan S, Wang P, Lin J, Zhou C, Zhou J (2023) Qwen-vl: A frontier large vision-language model with versatile abilities. arXiv preprint arXiv:230812966

\bibitem[{Bansal et~al.(2024)Bansal, Borgnia, Chu, Li, Kazemi, Huang, Goldblum, Geiping, and Goldstein}]{bansal2024cold}
Bansal A, Borgnia E, Chu HM, Li J, Kazemi H, Huang F, Goldblum M, Geiping J, Goldstein T (2024) Cold diffusion: Inverting arbitrary image transforms without noise. Advances in Neural Information Processing Systems 36

\bibitem[{Bao et~al.(2022)Bao, Li, Zhu, and Zhang}]{bao2022analytic}
Bao F, Li C, Zhu J, Zhang B (2022) Analytic-dpm: an analytic estimate of the optimal reverse variance in diffusion probabilistic models. arXiv preprint arXiv:220106503

\bibitem[{Bar-Tal et~al.(2022)Bar-Tal, Ofri-Amar, Fridman, Kasten, and Dekel}]{bar2022text2live}
Bar-Tal O, Ofri-Amar D, Fridman R, Kasten Y, Dekel T (2022) Text2live: Text-driven layered image and video editing. In: European conference on computer vision, Springer, pp 707--723

\bibitem[{Basu et~al.(2023)Basu, Saberi, Bhardwaj, Chegini, Massiceti, Sanjabi, Hu, and Feizi}]{basu2023editval}
Basu S, Saberi M, Bhardwaj S, Chegini AM, Massiceti D, Sanjabi M, Hu SX, Feizi S (2023) Editval: Benchmarking diffusion based text-guided image editing methods. arXiv preprint arXiv:231002426

\bibitem[{Bi{\'n}kowski et~al.(2018)Bi{\'n}kowski, Sutherland, Arbel, and Gretton}]{binkowski2018demystifying}
Bi{\'n}kowski M, Sutherland DJ, Arbel M, Gretton A (2018) Demystifying mmd gans. arXiv preprint arXiv:180101401

\bibitem[{Blattmann et~al.(2023)Blattmann, Rombach, Ling, Dockhorn, Kim, Fidler, and Kreis}]{blattmann2023align}
Blattmann A, Rombach R, Ling H, Dockhorn T, Kim SW, Fidler S, Kreis K (2023) Align your latents: High-resolution video synthesis with latent diffusion models. In: Proceedings of the IEEE/CVF Conference on Computer Vision and Pattern Recognition, pp 22563--22575

\bibitem[{Bochkovskiy et~al.(2020)Bochkovskiy, Wang, and Liao}]{bochkovskiy2020yolov4}
Bochkovskiy A, Wang CY, Liao HYM (2020) Yolov4: Optimal speed and accuracy of object detection. arXiv preprint arXiv:200410934

\bibitem[{Brack et~al.(2024)Brack, Friedrich, Kornmeier, Tsaban, Schramowski, Kersting, and Passos}]{brack2024ledits++}
Brack M, Friedrich F, Kornmeier K, Tsaban L, Schramowski P, Kersting K, Passos A (2024) Ledits++: Limitless image editing using text-to-image models. In: Proceedings of the IEEE/CVF Conference on Computer Vision and Pattern Recognition, pp 8861--8870

\bibitem[{Brock(2018)}]{brock2018large}
Brock A (2018) Large scale gan training for high fidelity natural image synthesis. arXiv preprint arXiv:180911096

\bibitem[{Brooks et~al.(2023)Brooks, Holynski, and Efros}]{brooks2023instructpix2pix}
Brooks T, Holynski A, Efros AA (2023) Instructpix2pix: Learning to follow image editing instructions. In: Proceedings of the IEEE/CVF Conference on Computer Vision and Pattern Recognition, pp 18392--18402

\bibitem[{Caesar et~al.(2018)Caesar, Uijlings, and Ferrari}]{caesar2018coco}
Caesar H, Uijlings J, Ferrari V (2018) Coco-stuff: Thing and stuff classes in context. In: Proceedings of the IEEE conference on computer vision and pattern recognition, pp 1209--1218

\bibitem[{Canny(1986)}]{canny1986computational}
Canny J (1986) A computational approach to edge detection. IEEE Transactions on pattern analysis and machine intelligence (6):679--698

\bibitem[{Cao et~al.(2024{\natexlab{a}})Cao, Tan, Gao, Xu, Chen, Heng, and Li}]{cao2024survey}
Cao H, Tan C, Gao Z, Xu Y, Chen G, Heng PA, Li SZ (2024{\natexlab{a}}) A survey on generative diffusion models. IEEE Transactions on Knowledge and Data Engineering

\bibitem[{Cao et~al.(2023)Cao, Wang, Qi, Shan, Qie, and Zheng}]{cao2023masactrl}
Cao M, Wang X, Qi Z, Shan Y, Qie X, Zheng Y (2023) Masactrl: Tuning-free mutual self-attention control for consistent image synthesis and editing. In: Proceedings of the IEEE/CVF International Conference on Computer Vision, pp 22560--22570

\bibitem[{Cao et~al.(2024{\natexlab{b}})Cao, Zhou, Song, and Yang}]{cao2024controllable}
Cao P, Zhou F, Song Q, Yang L (2024{\natexlab{b}}) Controllable generation with text-to-image diffusion models: A survey. arXiv preprint arXiv:240304279

\bibitem[{Cao et~al.(2017)Cao, Simon, Wei, and Sheikh}]{cao2017realtime}
Cao Z, Simon T, Wei SE, Sheikh Y (2017) Realtime multi-person 2d pose estimation using part affinity fields. In: Proceedings of the IEEE conference on computer vision and pattern recognition, pp 7291--7299

\bibitem[{Caron et~al.(2021)Caron, Touvron, Misra, J{\'e}gou, Mairal, Bojanowski, and Joulin}]{caron2021emerging}
Caron M, Touvron H, Misra I, J{\'e}gou H, Mairal J, Bojanowski P, Joulin A (2021) Emerging properties in self-supervised vision transformers. In: Proceedings of the IEEE/CVF international conference on computer vision, pp 9650--9660

\bibitem[{Chakrabarty et~al.(2023)Chakrabarty, Singh, Saakyan, and Muresan}]{chakrabarty2023learning}
Chakrabarty T, Singh K, Saakyan A, Muresan S (2023) Learning to follow object-centric image editing instructions faithfully. arXiv preprint arXiv:231019145

\bibitem[{Chen et~al.(2024{\natexlab{a}})Chen, Huang, Liu, Shen, Zhao, and Zhao}]{chen2024anydoor}
Chen X, Huang L, Liu Y, Shen Y, Zhao D, Zhao H (2024{\natexlab{a}}) Anydoor: Zero-shot object-level image customization. In: Proceedings of the IEEE/CVF Conference on Computer Vision and Pattern Recognition, pp 6593--6602

\bibitem[{Chen et~al.(2024{\natexlab{b}})Chen, Wu, Wang, Su, Chen, Xing, Zhong, Zhang, Zhu, Lu et~al.}]{chen2024internvl}
Chen Z, Wu J, Wang W, Su W, Chen G, Xing S, Zhong M, Zhang Q, Zhu X, Lu L, et~al. (2024{\natexlab{b}}) Internvl: Scaling up vision foundation models and aligning for generic visual-linguistic tasks. In: Proceedings of the IEEE/CVF conference on computer vision and pattern recognition, pp 24185--24198

\bibitem[{Choi et~al.(2021)Choi, Park, Lee, and Choo}]{choi2021viton}
Choi S, Park S, Lee M, Choo J (2021) Viton-hd: High-resolution virtual try-on via misalignment-aware normalization. In: Proceedings of the IEEE/CVF conference on computer vision and pattern recognition, pp 14131--14140

\bibitem[{Choi et~al.(2024)Choi, Kwak, Lee, Choi, and Shin}]{choi2024improving}
Choi Y, Kwak S, Lee K, Choi H, Shin J (2024) Improving diffusion models for virtual try-on. arXiv preprint arXiv:240305139

\bibitem[{Chu et~al.(2024)Chu, Qiao, Zhang, Xu, Wei, Yang, Sun, Hu, Lin, Zhang et~al.}]{chu2024mobilevlm}
Chu X, Qiao L, Zhang X, Xu S, Wei F, Yang Y, Sun X, Hu Y, Lin X, Zhang B, et~al. (2024) Mobilevlm v2: Faster and stronger baseline for vision language model. arXiv preprint arXiv:240203766

\bibitem[{Contributors(2020)}]{contributors2020openmmlab}
Contributors M (2020) Openmmlab pose estimation toolbox and benchmark

\bibitem[{Croitoru et~al.(2023)Croitoru, Hondru, Ionescu, and Shah}]{croitoru2023diffusion}
Croitoru FA, Hondru V, Ionescu RT, Shah M (2023) Diffusion models in vision: A survey. IEEE Transactions on Pattern Analysis and Machine Intelligence 45(9):10850--10869

\bibitem[{DeTone et~al.(2016)DeTone, Malisiewicz, and Rabinovich}]{detone2016deep}
DeTone D, Malisiewicz T, Rabinovich A (2016) Deep image homography estimation. arXiv preprint arXiv:160603798

\bibitem[{Dhariwal and Nichol(2021)}]{dhariwal2021diffusion}
Dhariwal P, Nichol A (2021) Diffusion models beat gans on image synthesis. Advances in neural information processing systems 34:8780--8794

\bibitem[{Dong et~al.(2023)Dong, Xue, Duan, and Han}]{dong2023prompt}
Dong W, Xue S, Duan X, Han S (2023) Prompt tuning inversion for text-driven image editing using diffusion models. In: Proceedings of the IEEE/CVF International Conference on Computer Vision, pp 7430--7440

\bibitem[{Esser et~al.(2021)Esser, Rombach, and Ommer}]{esser2021taming}
Esser P, Rombach R, Ommer B (2021) Taming transformers for high-resolution image synthesis. In: Proceedings of the IEEE/CVF conference on computer vision and pattern recognition, pp 12873--12883

\bibitem[{Fu et~al.(2023)Fu, Hu, Du, Wang, Yang, and Gan}]{fu2023guiding}
Fu TJ, Hu W, Du X, Wang WY, Yang Y, Gan Z (2023) Guiding instruction-based image editing via multimodal large language models. arXiv preprint arXiv:230917102

\bibitem[{Gal et~al.(2022)Gal, Alaluf, Atzmon, Patashnik, Bermano, Chechik, and Cohen-Or}]{gal2022image}
Gal R, Alaluf Y, Atzmon Y, Patashnik O, Bermano AH, Chechik G, Cohen-Or D (2022) An image is worth one word: Personalizing text-to-image generation using textual inversion. arXiv preprint arXiv:220801618

\bibitem[{Gao et~al.(2023)Gao, Liu, Zeng, Xu, Li, Luo, Liu, Zhen, and Zhang}]{gao2023implicit}
Gao S, Liu X, Zeng B, Xu S, Li Y, Luo X, Liu J, Zhen X, Zhang B (2023) Implicit diffusion models for continuous super-resolution. In: Proceedings of the IEEE/CVF conference on computer vision and pattern recognition, pp 10021--10030

\bibitem[{Ge et~al.(2023)Ge, Nah, Liu, Poon, Tao, Catanzaro, Jacobs, Huang, Liu, and Balaji}]{ge2023preserve}
Ge S, Nah S, Liu G, Poon T, Tao A, Catanzaro B, Jacobs D, Huang JB, Liu MY, Balaji Y (2023) Preserve your own correlation: A noise prior for video diffusion models. In: Proceedings of the IEEE/CVF International Conference on Computer Vision, pp 22930--22941

\bibitem[{Geng et~al.(2024)Geng, Yang, Hang, Li, Gu, Zhang, Bao, Zhang, Li, Hu et~al.}]{geng2024instructdiffusion}
Geng Z, Yang B, Hang T, Li C, Gu S, Zhang T, Bao J, Zhang Z, Li H, Hu H, et~al. (2024) Instructdiffusion: A generalist modeling interface for vision tasks. In: Proceedings of the IEEE/CVF Conference on Computer Vision and Pattern Recognition, pp 12709--12720

\bibitem[{Gong et~al.(2022)Gong, Li, Feng, Wu, and Kong}]{gong2022diffuseq}
Gong S, Li M, Feng J, Wu Z, Kong L (2022) Diffuseq: Sequence to sequence text generation with diffusion models. arXiv preprint arXiv:221008933

\bibitem[{Goodfellow et~al.(2014)Goodfellow, Pouget-Abadie, Mirza, Xu, Warde-Farley, Ozair, Courville, and Bengio}]{goodfellow2014generative}
Goodfellow I, Pouget-Abadie J, Mirza M, Xu B, Warde-Farley D, Ozair S, Courville A, Bengio Y (2014) Generative adversarial nets. Advances in neural information processing systems 27

\bibitem[{Gu et~al.(2022)Gu, Ko, Go, Lee, Lee, and Shin}]{gu2022towards}
Gu G, Ko B, Go S, Lee SH, Lee J, Shin M (2022) Towards light-weight and real-time line segment detection. In: Proceedings of the AAAI Conference on Artificial Intelligence, vol~36, pp 726--734

\bibitem[{Guo et~al.(2023{\natexlab{a}})Guo, Wang, Yang, Huang, Wang, Pfister, and Wen}]{guo2023shadowdiffusion}
Guo L, Wang C, Yang W, Huang S, Wang Y, Pfister H, Wen B (2023{\natexlab{a}}) Shadowdiffusion: When degradation prior meets diffusion model for shadow removal. In: Proceedings of the IEEE/CVF Conference on Computer Vision and Pattern Recognition, pp 14049--14058

\bibitem[{Guo et~al.(2023{\natexlab{b}})Guo, Xiao, Chang, Deng, and Yan}]{guo2023sky}
Guo Y, Xiao X, Chang Y, Deng S, Yan L (2023{\natexlab{b}}) From sky to the ground: A large-scale benchmark and simple baseline towards real rain removal. In: Proceedings of the IEEE/CVF International Conference on Computer Vision, pp 12097--12107

\bibitem[{Han et~al.(2024)Han, Wen, Chen, Zhang, Song, Ren, Gao, Stathopoulos, He, Chen et~al.}]{han2024proxedit}
Han L, Wen S, Chen Q, Zhang Z, Song K, Ren M, Gao R, Stathopoulos A, He X, Chen Y, et~al. (2024) Proxedit: Improving tuning-free real image editing with proximal guidance. In: Proceedings of the IEEE/CVF Winter Conference on Applications of Computer Vision, pp 4291--4301

\bibitem[{He et~al.(2024)He, Ma, Huang, Huang, Gao, Wei, Dai, Han, and Liu}]{he2024freeedit}
He R, Ma K, Huang L, Huang S, Gao J, Wei X, Dai J, Han J, Liu S (2024) Freeedit: Mask-free reference-based image editing with multi-modal instruction. arXiv preprint arXiv:240918071

\bibitem[{Hertz et~al.(2022)Hertz, Mokady, Tenenbaum, Aberman, Pritch, and Cohen-Or}]{hertz2022prompt}
Hertz A, Mokady R, Tenenbaum J, Aberman K, Pritch Y, Cohen-Or D (2022) Prompt-to-prompt image editing with cross attention control. arXiv preprint arXiv:220801626

\bibitem[{Hessel et~al.(2021)Hessel, Holtzman, Forbes, Bras, and Choi}]{hessel2021clipscore}
Hessel J, Holtzman A, Forbes M, Bras RL, Choi Y (2021) Clipscore: A reference-free evaluation metric for image captioning. arXiv preprint arXiv:210408718

\bibitem[{Heusel et~al.(2017)Heusel, Ramsauer, Unterthiner, Nessler, and Hochreiter}]{heusel2017gans}
Heusel M, Ramsauer H, Unterthiner T, Nessler B, Hochreiter S (2017) Gans trained by a two time-scale update rule converge to a local nash equilibrium. Advances in neural information processing systems 30

\bibitem[{Higgins et~al.(2017)Higgins, Matthey, Pal, Burgess, Glorot, Botvinick, Mohamed, and Lerchner}]{higgins2017beta}
Higgins I, Matthey L, Pal A, Burgess CP, Glorot X, Botvinick MM, Mohamed S, Lerchner A (2017) beta-vae: Learning basic visual concepts with a constrained variational framework. ICLR (Poster) 3

\bibitem[{Ho and Salimans(2022)}]{ho2022classifier}
Ho J, Salimans T (2022) Classifier-free diffusion guidance. arXiv preprint arXiv:220712598

\bibitem[{Ho et~al.(2020)Ho, Jain, and Abbeel}]{ho2020denoising}
Ho J, Jain A, Abbeel P (2020) Denoising diffusion probabilistic models. Advances in neural information processing systems 33:6840--6851

\bibitem[{Ho et~al.(2022{\natexlab{a}})Ho, Chan, Saharia, Whang, Gao, Gritsenko, Kingma, Poole, Norouzi, Fleet et~al.}]{ho2022imagen}
Ho J, Chan W, Saharia C, Whang J, Gao R, Gritsenko A, Kingma DP, Poole B, Norouzi M, Fleet DJ, et~al. (2022{\natexlab{a}}) Imagen video: High definition video generation with diffusion models. arXiv preprint arXiv:221002303

\bibitem[{Ho et~al.(2022{\natexlab{b}})Ho, Saharia, Chan, Fleet, Norouzi, and Salimans}]{ho2022cascaded}
Ho J, Saharia C, Chan W, Fleet DJ, Norouzi M, Salimans T (2022{\natexlab{b}}) Cascaded diffusion models for high fidelity image generation. Journal of Machine Learning Research 23(47):1--33

\bibitem[{Ho et~al.(2022{\natexlab{c}})Ho, Salimans, Gritsenko, Chan, Norouzi, and Fleet}]{ho2022video}
Ho J, Salimans T, Gritsenko A, Chan W, Norouzi M, Fleet DJ (2022{\natexlab{c}}) Video diffusion models. Advances in Neural Information Processing Systems 35:8633--8646

\bibitem[{Huang et~al.(2024{\natexlab{a}})Huang, Huang, Liu, Yan, Lv, Liu, Xiong, Zhang, Chen, and Cao}]{huang2024diffusion}
Huang Y, Huang J, Liu Y, Yan M, Lv J, Liu J, Xiong W, Zhang H, Chen S, Cao L (2024{\natexlab{a}}) Diffusion model-based image editing: A survey. arXiv preprint arXiv:240217525

\bibitem[{Huang et~al.(2024{\natexlab{b}})Huang, Xie, Wang, Yuan, Cun, Ge, Zhou, Dong, Huang, Zhang et~al.}]{huang2024smartedit}
Huang Y, Xie L, Wang X, Yuan Z, Cun X, Ge Y, Zhou J, Dong C, Huang R, Zhang R, et~al. (2024{\natexlab{b}}) Smartedit: Exploring complex instruction-based image editing with multimodal large language models. In: Proceedings of the IEEE/CVF Conference on Computer Vision and Pattern Recognition, pp 8362--8371

\bibitem[{Huberman-Spiegelglas et~al.(2024)Huberman-Spiegelglas, Kulikov, and Michaeli}]{huberman2024edit}
Huberman-Spiegelglas I, Kulikov V, Michaeli T (2024) An edit friendly ddpm noise space: Inversion and manipulations. In: Proceedings of the IEEE/CVF Conference on Computer Vision and Pattern Recognition, pp 12469--12478

\bibitem[{Hui et~al.(2024)Hui, Yang, Zhao, Shi, Wang, Wang, Zhou, and Xie}]{hui2024hq}
Hui M, Yang S, Zhao B, Shi Y, Wang H, Wang P, Zhou Y, Xie C (2024) Hq-edit: A high-quality dataset for instruction-based image editing. arXiv preprint arXiv:240409990

\bibitem[{Isola et~al.(2017)Isola, Zhu, Zhou, and Efros}]{isola2017image}
Isola P, Zhu JY, Zhou T, Efros AA (2017) Image-to-image translation with conditional adversarial networks. In: Proceedings of the IEEE conference on computer vision and pattern recognition, pp 1125--1134

\bibitem[{Karaev et~al.(2024)Karaev, Makarov, Wang, Neverova, Vedaldi, and Rupprecht}]{karaev2024cotracker3}
Karaev N, Makarov I, Wang J, Neverova N, Vedaldi A, Rupprecht C (2024) Cotracker3: Simpler and better point tracking by pseudo-labelling real videos. arXiv preprint arXiv:241011831

\bibitem[{Karras(2019)}]{karras2019style}
Karras T (2019) A style-based generator architecture for generative adversarial networks. arXiv preprint arXiv:181204948

\bibitem[{Kawar et~al.(2023)Kawar, Zada, Lang, Tov, Chang, Dekel, Mosseri, and Irani}]{kawar2023imagic}
Kawar B, Zada S, Lang O, Tov O, Chang H, Dekel T, Mosseri I, Irani M (2023) Imagic: Text-based real image editing with diffusion models. In: Proceedings of the IEEE/CVF Conference on Computer Vision and Pattern Recognition, pp 6007--6017

\bibitem[{Kim et~al.(2022)Kim, Kwon, and Ye}]{kim2022diffusionclip}
Kim G, Kwon T, Ye JC (2022) Diffusionclip: Text-guided diffusion models for robust image manipulation. In: Proceedings of the IEEE/CVF conference on computer vision and pattern recognition, pp 2426--2435

\bibitem[{Kingma(2013)}]{kingma2013auto}
Kingma DP (2013) Auto-encoding variational bayes. arXiv preprint arXiv:13126114

\bibitem[{Kirillov et~al.(2023)Kirillov, Mintun, Ravi, Mao, Rolland, Gustafson, Xiao, Whitehead, Berg, Lo et~al.}]{kirillov2023segment}
Kirillov A, Mintun E, Ravi N, Mao H, Rolland C, Gustafson L, Xiao T, Whitehead S, Berg AC, Lo WY, et~al. (2023) Segment anything. In: Proceedings of the IEEE/CVF International Conference on Computer Vision, pp 4015--4026

\bibitem[{Kolkin et~al.(2019)Kolkin, Salavon, and Shakhnarovich}]{kolkin2019style}
Kolkin N, Salavon J, Shakhnarovich G (2019) Style transfer by relaxed optimal transport and self-similarity. In: Proceedings of the IEEE/CVF conference on computer vision and pattern recognition, pp 10051--10060

\bibitem[{Kong et~al.(2020)Kong, Ping, Huang, Zhao, and Catanzaro}]{kong2020diffwave}
Kong Z, Ping W, Huang J, Zhao K, Catanzaro B (2020) Diffwave: A versatile diffusion model for audio synthesis. arXiv preprint arXiv:200909761

\bibitem[{Kreiss et~al.(2021)Kreiss, Bertoni, and Alahi}]{kreiss2021openpifpaf}
Kreiss S, Bertoni L, Alahi A (2021) Openpifpaf: Composite fields for semantic keypoint detection and spatio-temporal association. IEEE Transactions on Intelligent Transportation Systems 23(8):13498--13511

\bibitem[{Lasinger et~al.(2019)Lasinger, Ranftl, Schindler, and Koltun}]{lasinger2019towards}
Lasinger K, Ranftl R, Schindler K, Koltun V (2019) Towards robust monocular depth estimation: Mixing datasets for zero-shot cross-dataset transfer. arXiv preprint arXiv:190701341

\bibitem[{Lee et~al.(2024)Lee, Jung, Lee, and Lee}]{lee2024semanticdraw}
Lee J, Jung DS, Lee K, Lee KM (2024) Semanticdraw: towards real-time interactive content creation from image diffusion models. arXiv preprint arXiv:240309055

\bibitem[{Lee and Park(2020)}]{lee2020centermask}
Lee Y, Park J (2020) Centermask: Real-time anchor-free instance segmentation. In: Proceedings of the IEEE/CVF conference on computer vision and pattern recognition, pp 13906--13915

\bibitem[{Levin and Fried(2023)}]{levin2023differential}
Levin E, Fried O (2023) Differential diffusion: Giving each pixel its strength. arXiv preprint arXiv:230600950

\bibitem[{Li et~al.(2023{\natexlab{a}})Li, Li, and Hoi}]{li2023blip}
Li D, Li J, Hoi S (2023{\natexlab{a}}) Blip-diffusion: Pre-trained subject representation for controllable text-to-image generation and editing. Advances in Neural Information Processing Systems 36:30146--30166

\bibitem[{Li et~al.(2022)Li, Li, Xiong, and Hoi}]{li2022blip}
Li J, Li D, Xiong C, Hoi S (2022) Blip: Bootstrapping language-image pre-training for unified vision-language understanding and generation. In: International conference on machine learning, PMLR, pp 12888--12900

\bibitem[{Li et~al.(2025)Li, Yang, Kuang, Wu, Wang, Xiao, and Chen}]{li2025controlnet}
Li M, Yang T, Kuang H, Wu J, Wang Z, Xiao X, Chen C (2025) Controlnet$++$: Improving conditional controls with efficient consistency feedback. In: European Conference on Computer Vision, Springer, pp 129--147

\bibitem[{Li et~al.(2024)Li, Liu, Singh, Wang, Zhang, Plummer, and Lin}]{li2024unihuman}
Li N, Liu Q, Singh KK, Wang Y, Zhang J, Plummer BA, Lin Z (2024) Unihuman: A unified model for editing human images in the wild. In: Proceedings of the IEEE/CVF Conference on Computer Vision and Pattern Recognition, pp 2039--2048

\bibitem[{Li et~al.(2023{\natexlab{b}})Li, Chen, and Lu}]{li2023moecontroller}
Li S, Chen C, Lu H (2023{\natexlab{b}}) Moecontroller: Instruction-based arbitrary image manipulation with mixture-of-expert controllers. arXiv preprint arXiv:230904372

\bibitem[{Li et~al.(2023{\natexlab{c}})Li, Singh, and Grover}]{li2023instructany2pix}
Li S, Singh H, Grover A (2023{\natexlab{c}}) Instructany2pix: Flexible visual editing via multimodal instruction following. arXiv preprint arXiv:231206738

\bibitem[{Lin et~al.(2014)Lin, Maire, Belongie, Hays, Perona, Ramanan, Doll{\'a}r, and Zitnick}]{lin2014microsoft}
Lin TY, Maire M, Belongie S, Hays J, Perona P, Ramanan D, Doll{\'a}r P, Zitnick CL (2014) Microsoft coco: Common objects in context. In: Computer Vision--ECCV 2014: 13th European Conference, Zurich, Switzerland, September 6-12, 2014, Proceedings, Part V 13, Springer, pp 740--755

\bibitem[{Liu et~al.(2024)Liu, Zhang, Qiu, Huang, Lin, Zhao, Geng, Lin, Jin, Zhang et~al.}]{liu2024sphinx}
Liu D, Zhang R, Qiu L, Huang S, Lin W, Zhao S, Geng S, Lin Z, Jin P, Zhang K, et~al. (2024) Sphinx-x: Scaling data and parameters for a family of multi-modal large language models. arXiv preprint arXiv:240205935

\bibitem[{Liu et~al.(2023)Liu, Chen, Yuan, Mei, Liu, Mandic, Wang, and Plumbley}]{liu2023audioldm}
Liu H, Chen Z, Yuan Y, Mei X, Liu X, Mandic D, Wang W, Plumbley MD (2023) Audioldm: Text-to-audio generation with latent diffusion models. arXiv preprint arXiv:230112503

\bibitem[{Liu et~al.(2021)Liu, Zhu, Pei, Fu, Qin, Zhang, Wan, and Feng}]{liu2021synthetic}
Liu Y, Zhu L, Pei S, Fu H, Qin J, Zhang Q, Wan L, Feng W (2021) From synthetic to real: Image dehazing collaborating with unlabeled real data. In: Proceedings of the 29th ACM international conference on multimedia, pp 50--58

\bibitem[{Ma et~al.(2022)Ma, Xu, Sun, Yan, Zhang, and Ji}]{ma2022x}
Ma Y, Xu G, Sun X, Yan M, Zhang J, Ji R (2022) X-clip: End-to-end multi-grained contrastive learning for video-text retrieval. In: Proceedings of the 30th ACM International Conference on Multimedia, pp 638--647

\bibitem[{Ma et~al.(2024)Ma, Ji, Ye, Lin, Wang, Zheng, Zhou, Sun, and Ji}]{ma2024i2ebench}
Ma Y, Ji J, Ye K, Lin W, Wang Z, Zheng Y, Zhou Q, Sun X, Ji R (2024) I2ebench: A comprehensive benchmark for instruction-based image editing. arXiv preprint arXiv:240814180

\bibitem[{Meng et~al.(2021)Meng, He, Song, Song, Wu, Zhu, and Ermon}]{meng2021sdedit}
Meng C, He Y, Song Y, Song J, Wu J, Zhu JY, Ermon S (2021) Sdedit: Guided image synthesis and editing with stochastic differential equations. arXiv preprint arXiv:210801073

\bibitem[{Meng et~al.(2023)Meng, Rombach, Gao, Kingma, Ermon, Ho, and Salimans}]{meng2023distillation}
Meng C, Rombach R, Gao R, Kingma D, Ermon S, Ho J, Salimans T (2023) On distillation of guided diffusion models. In: Proceedings of the IEEE/CVF Conference on Computer Vision and Pattern Recognition, pp 14297--14306

\bibitem[{Miao et~al.(2022)Miao, Wang, Wu, Li, Zhang, Wei, and Yang}]{miao2022large}
Miao J, Wang X, Wu Y, Li W, Zhang X, Wei Y, Yang Y (2022) Large-scale video panoptic segmentation in the wild: A benchmark. In: Proceedings of the IEEE/CVF Conference on Computer Vision and Pattern Recognition, pp 21033--21043

\bibitem[{Mirza(2014)}]{mirza2014conditional}
Mirza M (2014) Conditional generative adversarial nets. arXiv preprint arXiv:14111784

\bibitem[{Miyake et~al.(2023)Miyake, Iohara, Saito, and Tanaka}]{miyake2023negative}
Miyake D, Iohara A, Saito Y, Tanaka T (2023) Negative-prompt inversion: Fast image inversion for editing with text-guided diffusion models. arXiv preprint arXiv:230516807

\bibitem[{Mokady et~al.(2023)Mokady, Hertz, Aberman, Pritch, and Cohen-Or}]{mokady2023null}
Mokady R, Hertz A, Aberman K, Pritch Y, Cohen-Or D (2023) Null-text inversion for editing real images using guided diffusion models. In: Proceedings of the IEEE/CVF Conference on Computer Vision and Pattern Recognition, pp 6038--6047

\bibitem[{Morelli et~al.(2023)Morelli, Baldrati, Cartella, Cornia, Bertini, and Cucchiara}]{morelli2023ladi}
Morelli D, Baldrati A, Cartella G, Cornia M, Bertini M, Cucchiara R (2023) Ladi-vton: Latent diffusion textual-inversion enhanced virtual try-on. In: Proceedings of the 31st ACM International Conference on Multimedia, pp 8580--8589

\bibitem[{Mou et~al.(2024)Mou, Wang, Xie, Wu, Zhang, Qi, and Shan}]{mou2024t2i}
Mou C, Wang X, Xie L, Wu Y, Zhang J, Qi Z, Shan Y (2024) T2i-adapter: Learning adapters to dig out more controllable ability for text-to-image diffusion models. In: Proceedings of the AAAI Conference on Artificial Intelligence, vol~38, pp 4296--4304

\bibitem[{Nah et~al.(2017)Nah, Hyun~Kim, and Mu~Lee}]{nah2017deep}
Nah S, Hyun~Kim T, Mu~Lee K (2017) Deep multi-scale convolutional neural network for dynamic scene deblurring. In: Proceedings of the IEEE conference on computer vision and pattern recognition, pp 3883--3891

\bibitem[{Nichol et~al.(2021)Nichol, Dhariwal, Ramesh, Shyam, Mishkin, McGrew, Sutskever, and Chen}]{nichol2021glide}
Nichol A, Dhariwal P, Ramesh A, Shyam P, Mishkin P, McGrew B, Sutskever I, Chen M (2021) Glide: Towards photorealistic image generation and editing with text-guided diffusion models. arXiv preprint arXiv:211210741

\bibitem[{{\"O}zdenizci and Legenstein(2023)}]{ozdenizci2023restoring}
{\"O}zdenizci O, Legenstein R (2023) Restoring vision in adverse weather conditions with patch-based denoising diffusion models. IEEE Transactions on Pattern Analysis and Machine Intelligence 45(8):10346--10357

\bibitem[{Pan et~al.(2023)Pan, Gherardi, Xie, and Huang}]{pan2023effective}
Pan Z, Gherardi R, Xie X, Huang S (2023) Effective real image editing with accelerated iterative diffusion inversion. In: Proceedings of the IEEE/CVF International Conference on Computer Vision, pp 15912--15921

\bibitem[{Papamakarios et~al.(2017)Papamakarios, Pavlakou, and Murray}]{papamakarios2017masked}
Papamakarios G, Pavlakou T, Murray I (2017) Masked autoregressive flow for density estimation. Advances in neural information processing systems 30

\bibitem[{Park et~al.(2020)Park, Zhu, Wang, Lu, Shechtman, Efros, and Zhang}]{park2020swapping}
Park T, Zhu JY, Wang O, Lu J, Shechtman E, Efros A, Zhang R (2020) Swapping autoencoder for deep image manipulation. Advances in Neural Information Processing Systems 33:7198--7211

\bibitem[{Parmar et~al.(2023)Parmar, Kumar~Singh, Zhang, Li, Lu, and Zhu}]{parmar2023zero}
Parmar G, Kumar~Singh K, Zhang R, Li Y, Lu J, Zhu JY (2023) Zero-shot image-to-image translation. In: ACM SIGGRAPH 2023 Conference Proceedings, pp 1--11

\bibitem[{Phung et~al.(2023)Phung, Dao, and Tran}]{phung2023wavelet}
Phung H, Dao Q, Tran A (2023) Wavelet diffusion models are fast and scalable image generators. In: Proceedings of the IEEE/CVF conference on computer vision and pattern recognition, pp 10199--10208

\bibitem[{Po et~al.(2024)Po, Yifan, Golyanik, Aberman, Barron, Bermano, Chan, Dekel, Holynski, Kanazawa et~al.}]{po2024state}
Po R, Yifan W, Golyanik V, Aberman K, Barron JT, Bermano A, Chan E, Dekel T, Holynski A, Kanazawa A, et~al. (2024) State of the art on diffusion models for visual computing. In: Computer Graphics Forum, Wiley Online Library, vol~43, p e15063

\bibitem[{Podell et~al.(2023)Podell, English, Lacey, Blattmann, Dockhorn, M{\"u}ller, Penna, and Rombach}]{podell2023sdxl}
Podell D, English Z, Lacey K, Blattmann A, Dockhorn T, M{\"u}ller J, Penna J, Rombach R (2023) Sdxl: Improving latent diffusion models for high-resolution image synthesis. arXiv preprint arXiv:230701952

\bibitem[{Qin et~al.(2023)Qin, Zhang, Yu, Feng, Yang, Zhou, Wang, Niebles, Xiong, Savarese et~al.}]{qin2023unicontrol}
Qin C, Zhang S, Yu N, Feng Y, Yang X, Zhou Y, Wang H, Niebles JC, Xiong C, Savarese S, et~al. (2023) Unicontrol: A unified diffusion model for controllable visual generation in the wild. arXiv preprint arXiv:230511147

\bibitem[{Radford et~al.(2021)Radford, Kim, Hallacy, Ramesh, Goh, Agarwal, Sastry, Askell, Mishkin, Clark et~al.}]{radford2021learning}
Radford A, Kim JW, Hallacy C, Ramesh A, Goh G, Agarwal S, Sastry G, Askell A, Mishkin P, Clark J, et~al. (2021) Learning transferable visual models from natural language supervision. In: International conference on machine learning, PMLR, pp 8748--8763

\bibitem[{Ramesh et~al.(2021)Ramesh, Pavlov, Goh, Gray, Voss, Radford, Chen, and Sutskever}]{ramesh2021zero}
Ramesh A, Pavlov M, Goh G, Gray S, Voss C, Radford A, Chen M, Sutskever I (2021) Zero-shot text-to-image generation. In: International conference on machine learning, Pmlr, pp 8821--8831

\bibitem[{Ramesh et~al.(2022)Ramesh, Dhariwal, Nichol, Chu, and Chen}]{ramesh2022hierarchical}
Ramesh A, Dhariwal P, Nichol A, Chu C, Chen M (2022) Hierarchical text-conditional image generation with clip latents. arXiv preprint arXiv:220406125 1(2):3

\bibitem[{Ranftl et~al.(2020)Ranftl, Lasinger, Hafner, Schindler, and Koltun}]{ranftl2020towards}
Ranftl R, Lasinger K, Hafner D, Schindler K, Koltun V (2020) Towards robust monocular depth estimation: Mixing datasets for zero-shot cross-dataset transfer. IEEE transactions on pattern analysis and machine intelligence 44(3):1623--1637

\bibitem[{Razavi et~al.(2019)Razavi, Van~den Oord, and Vinyals}]{razavi2019generating}
Razavi A, Van~den Oord A, Vinyals O (2019) Generating diverse high-fidelity images with vq-vae-2. Advances in neural information processing systems 32

\bibitem[{Rezatofighi et~al.(2019)Rezatofighi, Tsoi, Gwak, Sadeghian, Reid, and Savarese}]{rezatofighi2019generalized}
Rezatofighi H, Tsoi N, Gwak J, Sadeghian A, Reid I, Savarese S (2019) Generalized intersection over union: A metric and a loss for bounding box regression. In: Proceedings of the IEEE/CVF conference on computer vision and pattern recognition, pp 658--666

\bibitem[{Rezende and Mohamed(2015)}]{rezende2015variational}
Rezende D, Mohamed S (2015) Variational inference with normalizing flows. In: International conference on machine learning, PMLR, pp 1530--1538

\bibitem[{Rombach et~al.(2022)Rombach, Blattmann, Lorenz, Esser, and Ommer}]{rombach2022high}
Rombach R, Blattmann A, Lorenz D, Esser P, Ommer B (2022) High-resolution image synthesis with latent diffusion models. In: Proceedings of the IEEE/CVF conference on computer vision and pattern recognition, pp 10684--10695

\bibitem[{Rout et~al.(2024)Rout, Chen, Ruiz, Caramanis, Shakkottai, and Chu}]{rout2024semantic}
Rout L, Chen Y, Ruiz N, Caramanis C, Shakkottai S, Chu WS (2024) Semantic image inversion and editing using rectified stochastic differential equations. arXiv preprint arXiv:241010792

\bibitem[{Ruan et~al.(2023)Ruan, Ma, Yang, He, Liu, Fu, Yuan, Jin, and Guo}]{ruan2023mm}
Ruan L, Ma Y, Yang H, He H, Liu B, Fu J, Yuan NJ, Jin Q, Guo B (2023) Mm-diffusion: Learning multi-modal diffusion models for joint audio and video generation. In: Proceedings of the IEEE/CVF Conference on Computer Vision and Pattern Recognition, pp 10219--10228

\bibitem[{Ruan et~al.(2024)Ruan, Tian, Huang, Zhang, and Xiao}]{ruan2024univg}
Ruan L, Tian L, Huang C, Zhang X, Xiao X (2024) Univg: Towards unified-modal video generation. arXiv preprint arXiv:240109084

\bibitem[{Ruiz et~al.(2023)Ruiz, Li, Jampani, Pritch, Rubinstein, and Aberman}]{ruiz2023dreambooth}
Ruiz N, Li Y, Jampani V, Pritch Y, Rubinstein M, Aberman K (2023) Dreambooth: Fine tuning text-to-image diffusion models for subject-driven generation. In: Proceedings of the IEEE/CVF conference on computer vision and pattern recognition, pp 22500--22510

\bibitem[{Saharia et~al.(2022{\natexlab{a}})Saharia, Chan, Saxena, Li, Whang, Denton, Ghasemipour, Gontijo~Lopes, Karagol~Ayan, Salimans et~al.}]{saharia2022photorealistic}
Saharia C, Chan W, Saxena S, Li L, Whang J, Denton EL, Ghasemipour K, Gontijo~Lopes R, Karagol~Ayan B, Salimans T, et~al. (2022{\natexlab{a}}) Photorealistic text-to-image diffusion models with deep language understanding. Advances in neural information processing systems 35:36479--36494

\bibitem[{Saharia et~al.(2022{\natexlab{b}})Saharia, Ho, Chan, Salimans, Fleet, and Norouzi}]{saharia2022image}
Saharia C, Ho J, Chan W, Salimans T, Fleet DJ, Norouzi M (2022{\natexlab{b}}) Image super-resolution via iterative refinement. IEEE transactions on pattern analysis and machine intelligence 45(4):4713--4726

\bibitem[{Schuhmann et~al.(2022)Schuhmann, Beaumont, Vencu, Gordon, Wightman, Cherti, Coombes, Katta, Mullis, Wortsman et~al.}]{schuhmann2022laion}
Schuhmann C, Beaumont R, Vencu R, Gordon C, Wightman R, Cherti M, Coombes T, Katta A, Mullis C, Wortsman M, et~al. (2022) Laion-5b: An open large-scale dataset for training next generation image-text models. Advances in Neural Information Processing Systems 35:25278--25294

\bibitem[{Shang et~al.(2024)Shang, Shan, Liu, Wang, Wang, Zhang, and Zhang}]{shang2024resdiff}
Shang S, Shan Z, Liu G, Wang L, Wang X, Zhang Z, Zhang J (2024) Resdiff: Combining cnn and diffusion model for image super-resolution. In: Proceedings of the AAAI Conference on Artificial Intelligence, vol~38, pp 8975--8983

\bibitem[{Sheynin et~al.(2024)Sheynin, Polyak, Singer, Kirstain, Zohar, Ashual, Parikh, and Taigman}]{sheynin2024emu}
Sheynin S, Polyak A, Singer U, Kirstain Y, Zohar A, Ashual O, Parikh D, Taigman Y (2024) Emu edit: Precise image editing via recognition and generation tasks. In: Proceedings of the IEEE/CVF Conference on Computer Vision and Pattern Recognition, pp 8871--8879

\bibitem[{Shi et~al.(2024)Shi, Xue, Liew, Pan, Yan, Zhang, Tan, and Bai}]{shi2024dragdiffusion}
Shi Y, Xue C, Liew JH, Pan J, Yan H, Zhang W, Tan VY, Bai S (2024) Dragdiffusion: Harnessing diffusion models for interactive point-based image editing. In: Proceedings of the IEEE/CVF Conference on Computer Vision and Pattern Recognition, pp 8839--8849

\bibitem[{Shuai et~al.(2024)Shuai, Ding, Ma, Tu, Jiang, and Tao}]{shuai2024survey}
Shuai X, Ding H, Ma X, Tu R, Jiang YG, Tao D (2024) A survey of multimodal-guided image editing with text-to-image diffusion models. arXiv preprint arXiv:240614555

\bibitem[{Singer et~al.(2022)Singer, Polyak, Hayes, Yin, An, Zhang, Hu, Yang, Ashual, Gafni et~al.}]{singer2022make}
Singer U, Polyak A, Hayes T, Yin X, An J, Zhang S, Hu Q, Yang H, Ashual O, Gafni O, et~al. (2022) Make-a-video: Text-to-video generation without text-video data. arXiv preprint arXiv:220914792

\bibitem[{Sohl-Dickstein et~al.(2015)Sohl-Dickstein, Weiss, Maheswaranathan, and Ganguli}]{sohl2015deep}
Sohl-Dickstein J, Weiss E, Maheswaranathan N, Ganguli S (2015) Deep unsupervised learning using nonequilibrium thermodynamics. In: International conference on machine learning, PMLR, pp 2256--2265

\bibitem[{Song et~al.(2020{\natexlab{a}})Song, Meng, and Ermon}]{song2020denoising}
Song J, Meng C, Ermon S (2020{\natexlab{a}}) Denoising diffusion implicit models. arXiv preprint arXiv:201002502

\bibitem[{Song and Ermon(2019)}]{song2019generative}
Song Y, Ermon S (2019) Generative modeling by estimating gradients of the data distribution. Advances in neural information processing systems 32

\bibitem[{Song and Ermon(2020)}]{song2020improved}
Song Y, Ermon S (2020) Improved techniques for training score-based generative models. Advances in neural information processing systems 33:12438--12448

\bibitem[{Song et~al.(2020{\natexlab{b}})Song, Sohl-Dickstein, Kingma, Kumar, Ermon, and Poole}]{song2020score}
Song Y, Sohl-Dickstein J, Kingma DP, Kumar A, Ermon S, Poole B (2020{\natexlab{b}}) Score-based generative modeling through stochastic differential equations. arXiv preprint arXiv:201113456

\bibitem[{Song et~al.(2023)Song, Zhang, Lin, Cohen, Price, Zhang, Kim, and Aliaga}]{song2023objectstitch}
Song Y, Zhang Z, Lin Z, Cohen S, Price B, Zhang J, Kim SY, Aliaga D (2023) Objectstitch: Object compositing with diffusion model. In: Proceedings of the IEEE/CVF Conference on Computer Vision and Pattern Recognition, pp 18310--18319

\bibitem[{Su et~al.(2022)Su, Song, Meng, and Ermon}]{su2022dual}
Su X, Song J, Meng C, Ermon S (2022) Dual diffusion implicit bridges for image-to-image translation. arXiv preprint arXiv:220308382

\bibitem[{Su et~al.(2021)Su, Liu, Yu, Hu, Liao, Tian, Pietik{\"a}inen, and Liu}]{su2021pixel}
Su Z, Liu W, Yu Z, Hu D, Liao Q, Tian Q, Pietik{\"a}inen M, Liu L (2021) Pixel difference networks for efficient edge detection. In: Proceedings of the IEEE/CVF international conference on computer vision, pp 5117--5127

\bibitem[{Szegedy et~al.(2015)Szegedy, Liu, Jia, Sermanet, Reed, Anguelov, Erhan, Vanhoucke, and Rabinovich}]{szegedy2015going}
Szegedy C, Liu W, Jia Y, Sermanet P, Reed S, Anguelov D, Erhan D, Vanhoucke V, Rabinovich A (2015) Going deeper with convolutions. In: Proceedings of the IEEE conference on computer vision and pattern recognition, pp 1--9

\bibitem[{Tan et~al.(2024)Tan, Liu, Yang, Xue, and Wang}]{tan2024ominicontrol}
Tan Z, Liu S, Yang X, Xue Q, Wang X (2024) Ominicontrol: Minimal and universal control for diffusion transformer. arXiv preprint arXiv:241115098 3

\bibitem[{Team et~al.(2024)Team, Georgiev, Lei, Burnell, Bai, Gulati, Tanzer, Vincent, Pan, Wang et~al.}]{team2024gemini}
Team G, Georgiev P, Lei VI, Burnell R, Bai L, Gulati A, Tanzer G, Vincent D, Pan Z, Wang S, et~al. (2024) Gemini 1.5: Unlocking multimodal understanding across millions of tokens of context. arXiv preprint arXiv:240305530

\bibitem[{Tian et~al.(2025)Tian, Li, Xu, Yuan, Wang, and Shen}]{tian2025mige}
Tian X, Li W, Xu B, Yuan Y, Wang Y, Shen H (2025) Mige: A unified framework for multimodal instruction-based image generation and editing. arXiv preprint arXiv:250221291

\bibitem[{Tumanyan et~al.(2023)Tumanyan, Geyer, Bagon, and Dekel}]{tumanyan2023plug}
Tumanyan N, Geyer M, Bagon S, Dekel T (2023) Plug-and-play diffusion features for text-driven image-to-image translation. In: Proceedings of the IEEE/CVF Conference on Computer Vision and Pattern Recognition, pp 1921--1930

\bibitem[{Van Den~Oord et~al.(2016)Van Den~Oord, Kalchbrenner, and Kavukcuoglu}]{van2016pixel}
Van Den~Oord A, Kalchbrenner N, Kavukcuoglu K (2016) Pixel recurrent neural networks. In: International conference on machine learning, PMLR, pp 1747--1756

\bibitem[{Van Den~Oord et~al.(2017)Van Den~Oord, Vinyals et~al.}]{van2017neural}
Van Den~Oord A, Vinyals O, et~al. (2017) Neural discrete representation learning. Advances in neural information processing systems 30

\bibitem[{Vasiljevic et~al.(2019)Vasiljevic, Kolkin, Zhang, Luo, Wang, Dai, Daniele, Mostajabi, Basart, Walter et~al.}]{vasiljevic2019diode}
Vasiljevic I, Kolkin N, Zhang S, Luo R, Wang H, Dai FZ, Daniele AF, Mostajabi M, Basart S, Walter MR, et~al. (2019) Diode: A dense indoor and outdoor depth dataset. arXiv preprint arXiv:190800463

\bibitem[{Vinker et~al.(2023)Vinker, Alaluf, Cohen-Or, and Shamir}]{vinker2023clipascene}
Vinker Y, Alaluf Y, Cohen-Or D, Shamir A (2023) Clipascene: Scene sketching with different types and levels of abstraction. In: Proceedings of the IEEE/CVF International Conference on Computer Vision, pp 4146--4156

\bibitem[{Wallace et~al.(2023)Wallace, Gokul, and Naik}]{wallace2023edict}
Wallace B, Gokul A, Naik N (2023) Edict: Exact diffusion inversion via coupled transformations. In: Proceedings of the IEEE/CVF Conference on Computer Vision and Pattern Recognition, pp 22532--22541

\bibitem[{Wang et~al.(2023{\natexlab{a}})Wang, Zhang, Birsak, and Wonka}]{wang2023instructedit}
Wang Q, Zhang B, Birsak M, Wonka P (2023{\natexlab{a}}) Instructedit: Improving automatic masks for diffusion-based image editing with user instructions. arXiv preprint arXiv:230518047

\bibitem[{Wang et~al.(2023{\natexlab{b}})Wang, Saharia, Montgomery, Pont-Tuset, Noy, Pellegrini, Onoe, Laszlo, Fleet, Soricut et~al.}]{wang2023imagen}
Wang S, Saharia C, Montgomery C, Pont-Tuset J, Noy S, Pellegrini S, Onoe Y, Laszlo S, Fleet DJ, Soricut R, et~al. (2023{\natexlab{b}}) Imagen editor and editbench: Advancing and evaluating text-guided image inpainting. In: Proceedings of the IEEE/CVF conference on computer vision and pattern recognition, pp 18359--18369

\bibitem[{Wang et~al.(2021)Wang, Feiszli, Wang, and Tran}]{wang2021unidentified}
Wang W, Feiszli M, Wang H, Tran D (2021) Unidentified video objects: A benchmark for dense, open-world segmentation. In: Proceedings of the IEEE/CVF international conference on computer vision, pp 10776--10785

\bibitem[{Wang et~al.(2024)Wang, Lipson, and Deng}]{wang2024sea}
Wang Y, Lipson L, Deng J (2024) Sea-raft: Simple, efficient, accurate raft for optical flow. In: European Conference on Computer Vision, Springer, pp 36--54

\bibitem[{Wang et~al.(2003)Wang, Simoncelli, and Bovik}]{wang2003multiscale}
Wang Z, Simoncelli EP, Bovik AC (2003) Multiscale structural similarity for image quality assessment. In: The Thrity-Seventh Asilomar Conference on Signals, Systems \& Computers, 2003, Ieee, vol~2, pp 1398--1402

\bibitem[{Wei et~al.(2024)Wei, Xiong, Ren, Du, Zhang, and Chen}]{wei2024omniedit}
Wei C, Xiong Z, Ren W, Du X, Zhang G, Chen W (2024) Omniedit: Building image editing generalist models through specialist supervision. In: The Thirteenth International Conference on Learning Representations

\bibitem[{Wu and De~la Torre(2023)}]{wu2023latent}
Wu CH, De~la Torre F (2023) A latent space of stochastic diffusion models for zero-shot image editing and guidance. In: Proceedings of the IEEE/CVF International Conference on Computer Vision, pp 7378--7387

\bibitem[{Wu et~al.(2023)Wu, Ge, Wang, Lei, Gu, Shi, Hsu, Shan, Qie, and Shou}]{wu2023tune}
Wu JZ, Ge Y, Wang X, Lei SW, Gu Y, Shi Y, Hsu W, Shan Y, Qie X, Shou MZ (2023) Tune-a-video: One-shot tuning of image diffusion models for text-to-video generation. In: Proceedings of the IEEE/CVF International Conference on Computer Vision, pp 7623--7633

\bibitem[{Wu et~al.(2025)Wu, Huang, Wu, Cheng, Ding, and He}]{wu2025less}
Wu S, Huang M, Wu W, Cheng Y, Ding F, He Q (2025) Less-to-more generalization: Unlocking more controllability by in-context generation. arXiv preprint arXiv:250402160

\bibitem[{Xia et~al.(2023)Xia, Zhang, Wang, Wang, Wu, Tian, Yang, and Van~Gool}]{xia2023diffir}
Xia B, Zhang Y, Wang S, Wang Y, Wu X, Tian Y, Yang W, Van~Gool L (2023) Diffir: Efficient diffusion model for image restoration. In: Proceedings of the IEEE/CVF International Conference on Computer Vision, pp 13095--13105

\bibitem[{Xia et~al.(2024)Xia, Zhang, Li, Wang, Wang, Wu, Yu, and Jia}]{xia2024dreamomni}
Xia B, Zhang Y, Li J, Wang C, Wang Y, Wu X, Yu B, Jia J (2024) Dreamomni: Unified image generation and editing. arXiv preprint arXiv:241217098

\bibitem[{Xia et~al.(2025)Xia, Zhang, and Zhang}]{xia2025consistent}
Xia T, Zhang Y, Zhang TLL (2025) Consistent image layout editing with diffusion models. arXiv preprint arXiv:250306419

\bibitem[{Xiao et~al.(2024)Xiao, Wang, Zhou, Yuan, Xing, Yan, Wang, Huang, and Liu}]{xiao2024omnigen}
Xiao S, Wang Y, Zhou J, Yuan H, Xing X, Yan R, Wang S, Huang T, Liu Z (2024) Omnigen: Unified image generation. arXiv preprint arXiv:240911340

\bibitem[{Xie and Tu(2015)}]{xie2015holistically}
Xie S, Tu Z (2015) Holistically-nested edge detection. In: Proceedings of the IEEE international conference on computer vision, pp 1395--1403

\bibitem[{Xing et~al.(2024)Xing, Feng, Chen, Dai, Hu, Xu, Wu, and Jiang}]{xing2024survey}
Xing Z, Feng Q, Chen H, Dai Q, Hu H, Xu H, Wu Z, Jiang YG (2024) A survey on video diffusion models. ACM Computing Surveys 57(2):1--42

\bibitem[{Yang et~al.(2023{\natexlab{a}})Yang, Gu, Zhang, Zhang, Chen, Sun, Chen, and Wen}]{yang2023paint}
Yang B, Gu S, Zhang B, Zhang T, Chen X, Sun X, Chen D, Wen F (2023{\natexlab{a}}) Paint by example: Exemplar-based image editing with diffusion models. In: Proceedings of the IEEE/CVF Conference on Computer Vision and Pattern Recognition, pp 18381--18391

\bibitem[{Yang et~al.(2023{\natexlab{b}})Yang, Zhang, Song, Hong, Xu, Zhao, Zhang, Cui, and Yang}]{yang2023diffusion}
Yang L, Zhang Z, Song Y, Hong S, Xu R, Zhao Y, Zhang W, Cui B, Yang MH (2023{\natexlab{b}}) Diffusion models: A comprehensive survey of methods and applications. ACM Computing Surveys 56(4):1--39

\bibitem[{Ye et~al.(2023)Ye, Zhang, Liu, Han, and Yang}]{ye2023ip}
Ye H, Zhang J, Liu S, Han X, Yang W (2023) Ip-adapter: Text compatible image prompt adapter for text-to-image diffusion models. arXiv preprint arXiv:230806721

\bibitem[{Yu et~al.(2024)Yu, Chow, Yue, Pan, Wu, Wan, Li, Tang, Zhang, and Zhuang}]{yu2024anyedit}
Yu Q, Chow W, Yue Z, Pan K, Wu Y, Wan X, Li J, Tang S, Zhang H, Zhuang Y (2024) Anyedit: Mastering unified high-quality image editing for any idea. arXiv preprint arXiv:241115738

\bibitem[{Yu et~al.(2023)Yu, Xu, Zhang, Liu, Ye, Wu, Yan, Zhu, Xiong, Liang et~al.}]{yu2023mvimgnet}
Yu X, Xu M, Zhang Y, Liu H, Ye C, Wu Y, Yan Z, Zhu C, Xiong Z, Liang T, et~al. (2023) Mvimgnet: A large-scale dataset of multi-view images. In: Proceedings of the IEEE/CVF conference on computer vision and pattern recognition, pp 9150--9161

\bibitem[{Zeng et~al.(2024)Zeng, Song, Nie, Tian, Wang, and Liu}]{zeng2024cat}
Zeng J, Song D, Nie W, Tian H, Wang T, Liu AA (2024) Cat-dm: Controllable accelerated virtual try-on with diffusion model. In: Proceedings of the IEEE/CVF Conference on Computer Vision and Pattern Recognition, pp 8372--8382

\bibitem[{Zhan et~al.(2019)Zhan, Pan, Liu, Lin, and Loy}]{zhan2019self}
Zhan X, Pan X, Liu Z, Lin D, Loy CC (2019) Self-supervised learning via conditional motion propagation. In: Proceedings of the IEEE/CVF Conference on Computer Vision and Pattern Recognition, pp 1881--1889

\bibitem[{Zhan et~al.(2024)Zhan, Chen, Mei, Zhao, Chen, Chen, Lyu, and Wang}]{zhan2024conditional}
Zhan Z, Chen D, Mei JP, Zhao Z, Chen J, Chen C, Lyu S, Wang C (2024) Conditional image synthesis with diffusion models: A survey. arXiv preprint arXiv:240919365

\bibitem[{Zhang et~al.(2024{\natexlab{a}})Zhang, Mo, Chen, Sun, and Su}]{zhang2024magicbrush}
Zhang K, Mo L, Chen W, Sun H, Su Y (2024{\natexlab{a}}) Magicbrush: A manually annotated dataset for instruction-guided image editing. Advances in Neural Information Processing Systems 36

\bibitem[{Zhang et~al.(2023)Zhang, Rao, and Agrawala}]{zhang2023adding}
Zhang L, Rao A, Agrawala M (2023) Adding conditional control to text-to-image diffusion models. In: Proceedings of the IEEE/CVF International Conference on Computer Vision, pp 3836--3847

\bibitem[{Zhang et~al.(2018)Zhang, Isola, Efros, Shechtman, and Wang}]{zhang2018unreasonable}
Zhang R, Isola P, Efros AA, Shechtman E, Wang O (2018) The unreasonable effectiveness of deep features as a perceptual metric. In: Proceedings of the IEEE conference on computer vision and pattern recognition, pp 586--595

\bibitem[{Zhang et~al.(2024{\natexlab{b}})Zhang, Yang, Feng, Qin, Chen, Yu, Chen, Wang, Savarese, Ermon et~al.}]{zhang2024hive}
Zhang S, Yang X, Feng Y, Qin C, Chen CC, Yu N, Chen Z, Wang H, Savarese S, Ermon S, et~al. (2024{\natexlab{b}}) Hive: Harnessing human feedback for instructional visual editing. In: Proceedings of the IEEE/CVF Conference on Computer Vision and Pattern Recognition, pp 9026--9036

\bibitem[{Zhang et~al.(2025)Zhang, Zhou, Zeng, Xu, Li, and Zuo}]{zhang2025framepainter}
Zhang Y, Zhou X, Zeng Y, Xu H, Li H, Zuo W (2025) Framepainter: Endowing interactive image editing with video diffusion priors. arXiv preprint arXiv:250108225

\bibitem[{Zhao(2016)}]{zhao2016energy}
Zhao J (2016) Energy-based generative adversarial network. arXiv preprint arXiv:160903126

\bibitem[{Zhao et~al.(2024{\natexlab{a}})Zhao, Chen, Chen, Bao, Hao, Yuan, and Wong}]{zhao2024uni}
Zhao S, Chen D, Chen YC, Bao J, Hao S, Yuan L, Wong KYK (2024{\natexlab{a}}) Uni-controlnet: All-in-one control to text-to-image diffusion models. Advances in Neural Information Processing Systems 36

\bibitem[{Zhao et~al.(2024{\natexlab{b}})Zhao, Guan, Fan, Xu, Lin, Pan, and Feng}]{zhao2024fastdrag}
Zhao X, Guan J, Fan C, Xu D, Lin Y, Pan H, Feng P (2024{\natexlab{b}}) Fastdrag: Manipulate anything in one step. arXiv preprint arXiv:240515769

\bibitem[{Zhou et~al.(2017{\natexlab{a}})Zhou, Lapedriza, Khosla, Oliva, and Torralba}]{zhou2017places}
Zhou B, Lapedriza A, Khosla A, Oliva A, Torralba A (2017{\natexlab{a}}) Places: A 10 million image database for scene recognition. IEEE transactions on pattern analysis and machine intelligence 40(6):1452--1464

\bibitem[{Zhou et~al.(2017{\natexlab{b}})Zhou, Zhao, Puig, Fidler, Barriuso, and Torralba}]{zhou2017scene}
Zhou B, Zhao H, Puig X, Fidler S, Barriuso A, Torralba A (2017{\natexlab{b}}) Scene parsing through ade20k dataset. In: Proceedings of the IEEE conference on computer vision and pattern recognition, pp 633--641

\bibitem[{Zhou et~al.(2022)Zhou, Wang, Yan, Lv, Zhu, and Feng}]{zhou2022magicvideo}
Zhou D, Wang W, Yan H, Lv W, Zhu Y, Feng J (2022) Magicvideo: Efficient video generation with latent diffusion models. arXiv preprint arXiv:221111018

\bibitem[{Zhu et~al.(2017)Zhu, Park, Isola, and Efros}]{zhu2017unpaired}
Zhu JY, Park T, Isola P, Efros AA (2017) Unpaired image-to-image translation using cycle-consistent adversarial networks. In: Proceedings of the IEEE international conference on computer vision, pp 2223--2232

\end{thebibliography}

\end{document}